\documentclass[preprint2]{aastex61}
\usepackage{natbib,graphicx,longtable,afterpage,amsmath}

\shorttitle{AGNs in the Chandra-COSMOS Legacy Survey}
\shortauthors{Suh et al.}

\begin{document}

\title{Multi-wavelength properties of Type 1 and Type 2 AGN host galaxies in the Chandra-COSMOS Legacy Survey}

\author{Hyewon Suh}
\altaffiliation{Subaru Fellow}
\email{suh@naoj.org}
\affil{Subaru Telescope, National Astronomical Observatory of Japan (NAOJ), National Institutes of Natural Sciences (NINS), 650 North A'ohoku place, Hilo, HI 96720, USA}

\author{Francesca Civano}
\affiliation{Harvard-Smithsonian Center for Astrophysics, Cambridge, MA 02138, USA}

\author{G\"unther Hasinger}
\affiliation{European Space Astronomy Centre (ESA/ESAC), Director of Science, E-28691 Villanueva de la Ca\~{n}ada, Madrid, Spain}

\author{Elisabeta Lusso}
\affiliation{Centre for Extragalactic Astronomy, Department of Physics, Durham University, South Road, Durham DH1 3LE, UK}

\author{Stefano Marchesi}
\affiliation{Department of Physics \& Astronomy, Clemson University, Clemson, SC 29634, USA}

\author{Andreas Schulze}
\altaffiliation{EACOA Fellow}
\affil{National Astronomical Observatory of Japan, Mitaka, Tokyo 181-8588, Japan}

\author{Masato Onodera}
\affil{Subaru Telescope, National Astronomical Observatory of Japan (NAOJ), National Institutes of Natural Sciences (NINS), 650 North A'ohoku place, Hilo, HI 96720, USA}
\affil{Department of Astronomical Science, SOKENDAI (The Graduate University for Advanced Studies), Osawa 2-21-1, Mitaka, Tokyo, 181-8588, Japan}

\author{David J. Rosario}
\affiliation{Centre for Extragalactic Astronomy, Department of Physics, Durham University, South Road, Durham DH1 3LE, UK}

\author{David B. Sanders}
\affil{Institute for Astronomy, University of Hawaii, 2680 Woodlawn Drive, Honolulu, HI 96822, USA}

\begin{abstract}
We investigate the multi-wavelength properties of host galaxies of 3701 X-ray-selected active galactic nuclei (AGNs) out to $z\sim5$ in the {\it Chandra}-COSMOS Legacy Survey. Thanks to the extensive multi-wavelength photometry available in the COSMOS field, we derive AGN luminosities, host stellar masses, and star formation rates (SFRs) via a multi-component SED fitting technique. Type 1 and Type 2 AGNs follow the same intrinsic $L_{\rm 2-10keV}-L_{\rm 6\mu m}$ relation, suggesting that mid-infrared emission is a reasonably good measure of the AGN accretion power regardless of obscuration. We find that there is a strong increase in Type 1 AGN fraction toward higher AGN luminosity, possibly due to the fact that Type 1 AGNs tend to be hosted by more massive galaxies. The AGN luminosity and SFR are consistent with an increase toward high stellar mass, while both the M$_{\rm stellar}$-dependence is weaker towards the high-mass end, which could be interpreted as a consequence of quenching both star formation and AGN activity in massive galaxies. AGN host galaxies tend to have SFRs that are consistent with normal star-forming galaxies, independent of AGN luminosities. We confirm that black hole accretion rate and SFR are correlated up to $z\sim5$, when forming stars. The majority ($\sim$73\%) of our AGN sample are faint in the far-infrared, implying that the moderate-luminosity AGNs seem to be still active after the star formation is suppressed. It is not certain whether AGN activity plays a role in quenching the star formation. We conclude that both AGN activity and star formation might be more fundamentally related to host stellar mass. 
\end{abstract}
\keywords{black hole physics -- galaxies: active -- galaxies: nuclei -- galaxies: evolution -- quasars: general}

\section{Introduction}

A number of observations have shown that the growth of supermassive black holes (SMBHs) is tightly linked with their host galaxies, as revealed by correlations between the black hole mass and host galaxy properties, i.e., the $M_{\rm BH}-M_{\rm bulge}$ relation (e.g., \citealt{Kormendy95, Magorrian98, Haring04, Kormendy13, McConnell13}) and $M_{\rm BH}-\sigma$ relation (e.g., \citealt{Ferrarese00, Gebhardt00, Merritt01, Tremaine02, Gultekin09, Graham11, Schulze11, McConnell13, Woo13}). The growth of active galactic nuclei (AGNs) and the star formation history (SFH) show a remarkably similar evolutionary behavior through cosmic time, indicating that there is a broad connection between nuclear activity and star formation (e.g., \citealt{Madau96, Giacconi02, Cowie03, Steffen03, Ueda03, Barger05, Hasinger05, Hopkins07, Aird15}). While most theoretical models of galaxy evolution require an AGN as a mechanism to regulate the star formation (e.g., \citealt{Silk98, DiMatteo05, Hopkins06}), our current understanding of the impact of AGN on star formation is still under debate (see \citealt{Alexander12, Kormendy13, Heckman14} for recent reviews). 

In order to understand the impact of AGN activity on the evolution of galaxies, there have been a number of studies investigating the star formation properties of AGN host galaxies (e.g., \citealt{Lutz10, Shao10, Lusso11, Mainieri11, Harrison12, Mullaney12, Rovilos12, Santini12, Rosario13, Lanzuisi15, Suh17}). However, the conclusions have been widely controversial. Some studies showed suppressed star formation for the most luminous AGNs (e.g., \citealt{Page12, Barger15}), whereas some others reported enhanced star formation in AGN host galaxies (e.g., \citealt{Lutz10, Mullaney12, Rovilos12, Santini12}). On the other hand, there are also studies presenting that the star formation as being independent of AGN activity, especially for moderate-luminosity AGNs (e.g., \citealt{Shao10, Mainieri11, Harrison12, Rosario12, Azadi15, Stanley15, Suh17}). The conflicting results for star formation in AGN host galaxies can be partially attributed to the different nature of the samples (i.e., small number statistics, selection biases) as well as the use of various methods to measure the parameters (i.e., use of different star formation rate (SFR) indicator and/or AGN luminosity). The sample selection including completeness and biases due to a specific selection method could introduce systematics: for example, infrared-selected AGNs may be biased toward higher SFR (e.g., \citealt{Chang17}), and the most massive black holes may be hosted by the most massive galaxies. Also, the most luminous AGNs might not represent the general AGN population, because they are a rare subset of all accreting black holes. Therefore, the underlying correlation between SFR, stellar mass, and redshift should be accounted for when studying the star formation in AGN host galaxies.

X-ray surveys are practically efficient for selecting AGNs because the X-ray emission is a relatively clean signal from the nuclear component that is produced within a few gravitational radii from the central accreting disk (e.g., \citealt{DeMarco13, Kara15}). X-ray-selected AGNs are less affected by obscuration, and also the contamination from non-nuclear emission, mainly due to star-formation processes, is far less significant than in optical and infrared surveys \citep{Donley08, Donley12, Lehmer12, Stern12}. The deep, large-area surveys observed by {\it Chandra} (i.e., {\it Chandra}-COSMOS Survey, \citealt{Elvis09}; {\it Chandra}-COSMOS Legacy Survey, \citealt{Civano16}) allow us to study a fairly large sample of AGNs over a broad range of luminosities ($41<\log~L_{\rm 0.5-10~keV}~{\rm erg~s^{-1}}<45$) out to $z\sim5$, providing a unique opportunity to study the evolution of black holes and galaxies. Furthermore, soft X-ray emission is partially absorbed by the hot dust surrounding the central black hole and re-emitted in the infrared, providing a crucial information on the structure and physical properties of the nuclear region (i.e., torus). Several studies have found a strong correlation between X-ray and mid-infrared (MIR) luminosities (e.g., \citealt{Lutz04, Fiore09, Gandhi09, Lanzuisi09, Lusso11, Asmus15, Stern15}), for which the MIR luminosity has also been used as a robust indicator of an intrinsic AGN power. 

According to the classical simplest AGN unification model, the observed classification of Type 1 and Type 2 AGNs, which depends on the presence of broad emission lines in their optical spectra, can be explained by the orientation effect of the dusty torus and anisotropic obscuration (e.g., \citealt{Antonucci93, Urry95, Netzer15}). However, several studies have reported challenges to this orientation-based scheme in that the fraction of Type 2 AGNs shows a clear anticorrelation with AGN luminosity (see, e.g., \citealt{Ueda03, Ueda14, Hasinger08, Lusso13, Merloni14, Aird15}), suggesting that there might be an intrinsic difference between obscured and unobscured AGNs. Recent studies suggested that the obscuration of AGNs is driven by the SMBH accretion properties (e.g., Eddington ratios; \citealt{Ricci17}) or the dust located in the host galaxy \citep{Goulding12}. Furthermore, there have been studies suggesting that the nuclear dust is not uniformly distributed around the central engine, indicating the complex and clumpy structure of dusty torus (e.g., \citealt{Ramos09, Ramos11, Markowitz14, Ichikawa15}). On the other hand, \citet{Sanders88} suggested an evolutionary scenario for AGNs (see also \citealt{DiMatteo05, Hopkins06b}) in which the obscuration is possibly a particular evolutionary phase, which is triggered by an accretion event (i.e., merger). In this evolutionary scheme, the obscured AGNs expel most of the obscuring material via AGN feedback, and evolve to an unobscured phase (i.e, Type 1 AGN) while consuming the remaining gas.

In this paper, we investigate the properties of Type 1 and Type 2 AGN host galaxies in the {\it Chandra}-COSMOS Legacy Survey (CCLS; \citealt{Civano16}) by exploiting a large sample of X-ray-selected moderate-luminosity AGNs to have a better understanding of the nuclear activity and its connection to the star formation. In this analysis, we consider 3701 X-ray-selected AGNs (985 Type 1 and 2716 Type 2 AGNs) in the CCLS, and analyze their multi-wavelength properties. Thanks to the large, uniform X-ray depth and the excellent extensive multi-wavelength data in the COSMOS field, we estimate the properties of both AGNs and their host galaxies in a wide range of redshifts for the largest data set adopted so far in this kind of study. We utilize multi-wavelength data from near-ultraviolet (NUV) to far-infrared (FIR) wavelengths and develop a multi-component spectral energy distribution (SED) fitting technique to decompose the SED into separate components. We provide the results and discuss the effects of the nuclear activity on the star formation in both Type 1 and Type 2 AGN host galaxies. 

Throughout this paper we assume a $\Lambda$CDM cosmology with $\Omega_{m}=0.3,~\Omega_{\Lambda}=0.7$, and $H_{0}=70~{\rm km~s^{-1}~Mpc^{-1}}$.

\section{AGN Sample}

We select a sample of AGNs from the CCLS catalog \citep{Civano16}, which comprises 4016 X-ray point sources detected by {\it Chandra} over a large area of $\sim2.2$ deg$^{2}$ in the COSMOS field. All the details about the full catalog of CCLS have been presented by \citet{Civano16} and \citet{Marchesi16}, including X-ray and optical/infrared photometric and spectroscopic properties. We consider 3701 X-ray selected AGNs, which have a reliable optical counterpart and spectroscopic and/or photometric redshift as in \citet{Marchesi16}. The spectroscopic information is available for $\sim$45\% (1665) of the sources, while for $\sim$55\% (2036) of the sources, only photometric redshifts are available. The photometric redshifts have been obtained using the publicly available code LePhare (e.g., \citealt{Arnouts99, Ilbert06, Salvato11}). The sources with $i_{\rm AB}<22.5$ mag have an uncertainty in the photometric redshift of $\sim$0.012, while sources fainter than $i_{\rm AB}=22.5$ mag have an uncertainty of $\sim$0.033. 

We take the spectroscopic and photometric classifications of the sources, which are described in detail in \citet[see Table 7]{Marchesi16}. From the catalog, 44\% of sources have information on spectral type. Of these sources, 632 (36\%) show evidence of at least one broad line in their spectra, while 1049 (59\%) show only narrow emission lines and/or absorption lines. While $\sim$56\% of sources are still without spectroscopic type, $\sim$96\% of the sample have a photometric SED template information. Approximately 23\% of sources are fitted with an unobscured AGN template, $\sim$9\% are fitted with an obscured AGN template, and $\sim$64\% by a template with an inactive galaxy. About 82\% of the sources with broad lines in their spectra have been fitted with an unobscured AGN template, while $\sim$97\% of the non-broad-line sources are fitted with either a galaxy template (74\%) or with an obscured AGN template (23\%). 

Finally, 1034 sources are classified as broad-line and/or unobscured AGN (hereafter, ``Type 1" AGN) from their optical spectrum, i.e., broad emission lines with FWHM larger than 2000 km/s, or their photometric SED is best fitted by an unobscured AGN template. Within the main sample, 2716 sources are classified as non-broad-line and/or obscured AGN (hereafter, ``Type 2" AGN), i.e., they show only narrow emission-line and/or absorption-line features in their spectra, or their photometric SED is best fitted by an obscured AGN template or a galaxy template.

\subsection{Photometric data}

We compile the SEDs of our sample of Type 1 and Type 2 AGN host galaxies from NUV (2300\AA) to FIR (500$\mu$m) wavelengths using the most recent multi-wavelength photometric catalog of the COSMOS field from \citet{Laigle16}. The catalog includes the {\it GALEX} NUV band, CFHT $U$ band, five Subaru Suprime-Cam bands ($B, V, r, i, z^{+}$), four UltraVista bands ({\it Y, H, J, Ks}), and four {\it Spitzer}/Infrared Array Camera (IRAC) bands (3.6, 4.5, 5.8, and 8.0$\mu$m). In addition, we use the 24 and 70$\mu$m bands of the Multiband Imaging Photometer for {\it Spitzer} (MIPS, \citealt{Sanders07, LeFloch09}) with $\sim$63\% (2317/3701) of the sources detected in the 24$\mu$m band, which is particularly important for identifying the AGN dusty obscuring structure. We also constrain the SEDs in the FIR wavelength range for $\sim$27\% (1011/3701) of the sources that have been detected by the {\it Herschel Space Observatory} (\citealt{Griffin10, Pilbratt10, Poglitsch10}: PACS 100$\mu$m ($\sim$15\%; 543/3701), 160$\mu$m ($\sim$12\%; 457/3701) and SPIRE 250$\mu$m ($\sim$22\%; 798/3701), 350$\mu$m ($\sim$11\%; 409/3701), 500$\mu$m ($\sim$3\%; 112/3701)).  

\section{SED Fitting}

The emission from the nuclear accretion disk peaks in the UV, and is partially absorbed by the dust and re-emitted in the IR wavelength range. The observed SEDs of AGNs thus present two peaks: one in the UV and another at the MIR wavelengths (e.g., \citealt{Elvis94, Richards06, Elvis12}). We use model templates including UV-optical emission from the AGN accretion disk around the SMBH, i.e., ``big blue bump" (BBB, \citealt{Sanders89, Elvis94, Elvis12, Richards06, Shang11, Krawczyk13}), dust emission from an AGN torus, galaxy emission from stellar populations, and FIR emission from a starburst to match the broadband photometric SEDs of the AGN sample. The nuclear emission of Type 1 AGNs contributes significantly to the UV-optical parts of the spectra (e.g., \citealt{Elvis12, Hao13}). On the other hand, for Type 2 AGNs, the nuclear emission dominates the SED only in the X-ray, and at other wavelengths the light is mainly due to the galaxy emission combined with reprocessed nuclear emission in the IR (e.g., \citealt{Lusso13, Suh17}). While nuclear emission, reprocessed by dust, could significantly contribute to the MIR luminosity, the FIR luminosity is known to be dominated by galaxy emission produced by star formation (e.g., \citealt{Kirkpatrick12}). 

In our custom SED fitting code, we have considered the same SED libraries as in AGNfitter\footnote{https://github.com/GabrielaCR/AGNfitter} \citep{Calistro16} for the different components of the observed SED, specifically: the FIR cold dust, the torus, the stellar population, and the accretion disk. We briefly summarize the main features of these libraries below. 

\begin{figure*}
\centering
\includegraphics[width=1\textwidth]{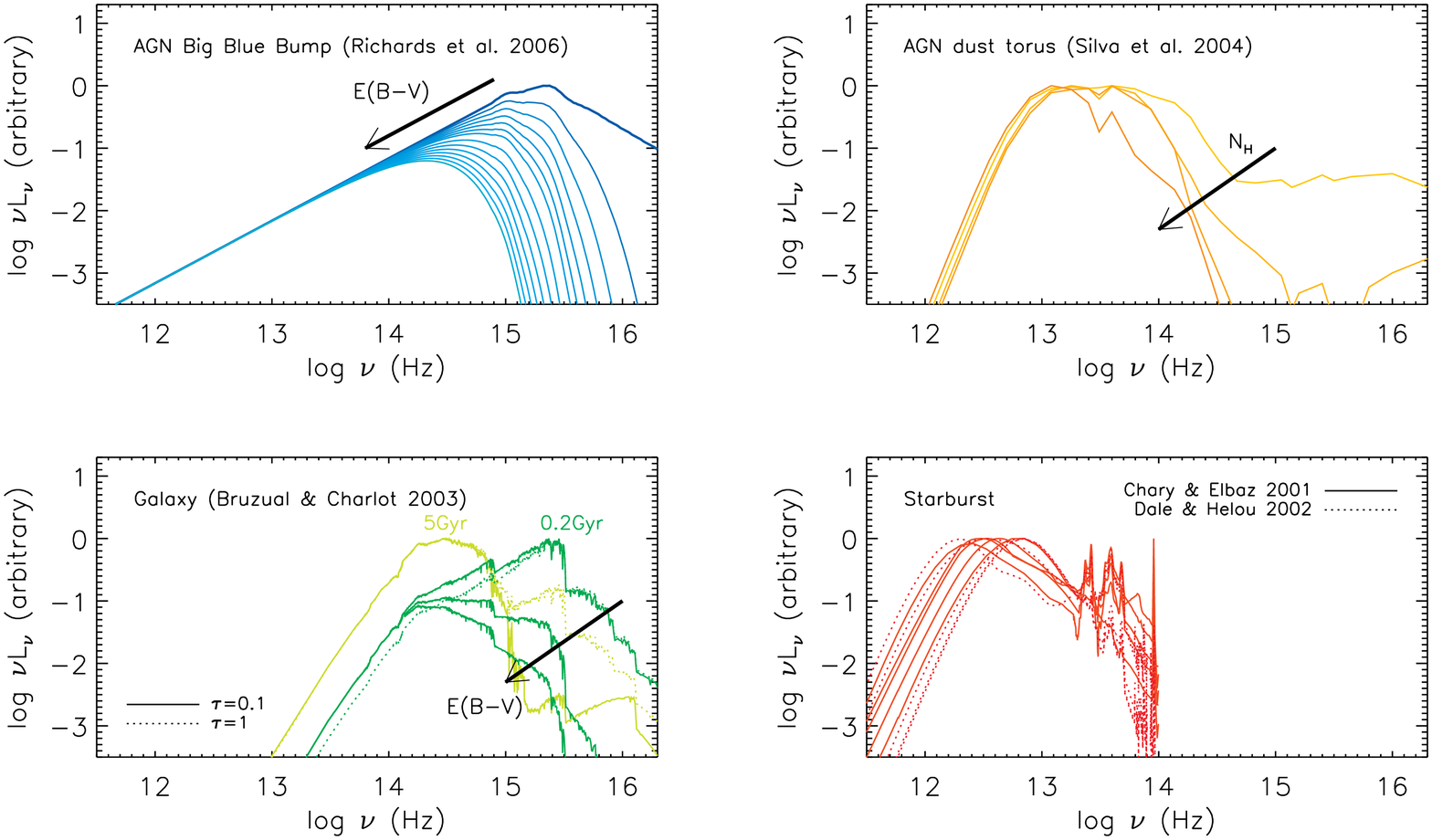}
\caption{Examples of model templates used in the multi-component SED fitting (see also \citealt{Lusso13, Calistro16}). Top left: blue curves indicate subsamples of AGN BBB templates \citep{Richards06} with different reddening levels $E(B-V)=0.0-1.0$. Top right: yellow curves correspond to four AGN dust torus templates \citep{Silva04} depending on the hydrogen column density, $N_{\rm H}$. Bottom left: green curves indicate some examples of host galaxy templates \citep{BC03} with various combinations of $\tau$=[0.1, 1] and $t_{\rm age}$=[0.2 Gyr, 5 Gyr] with $E(B-V)$=[0.0, 0.3, 0.5]. Bottom right: red curves correspond to the subset of starburst templates \citep{Chary01, Dale02}.}
\label{fig:templates}
\end{figure*}

\subsection{Model Templates}

The SED fitting technique used in this paper is a modified version of that in \citet{Suh17}, which is applied to Type 2 AGN host galaxies. For Type 2 AGN host galaxies, we decompose the SED into a nuclear AGN torus, a host galaxy with stellar populations, and a starburst component. A full detailed description of SED fitting for Type 2 AGN host galaxies in the CCLS is presented in \citet{Suh17}. For Type 1 AGN host galaxies, we add an additional fourth component in the fit, which is an AGN BBB template in the UV-optical range, taken from the mean quasar SED of \citet{Richards06}. This template is reddened according to the reddening law of \citet{Prevot84} for the Small Magellanic Cloud (SMC), which is found to be effective in treating the reddening in Type 1 AGNs (e.g., \citealt{Hopkins04, Salvato09}). The $E(B-V)_{\rm AGN}$ values range between 0 and 1 with a variable step ($\Delta E(B-V)_{\rm AGN}=0.01$ for $E(B-V)_{\rm AGN}$ between 0 and 0.1, and $\Delta E(B-V)_{\rm AGN}=0.05$ for $E(B-V)_{\rm AGN}$ between 0.1 and 1) for a total of 29 templates. A subsample of BBB templates with different reddening levels is presented in the top left panel of Figure~\ref{fig:templates} (blue curves).

The dust torus SED templates are taken from \citet{Silva04}, as constructed from the study of a large sample of Seyfert galaxies for which clear signatures of non-stellar nuclear emission were detected in the NIR and MIR, and also using the radiative transfer code GRASIL \citep{Silva98}. There are four different templates depending on the amount of nuclear obscuration in terms of hydrogen column density, ${\rm N_{H} < 10^{22}~cm^{-2}}$ for Seyfert 1, and ${\rm 10^{22} < N_{H} < 10^{23}~cm^{-2}}$, ${\rm 10^{23} < N_{H} < 10^{24}~cm^{-2}}$, and ${\rm N_{H} > 10^{24}~cm^{-2}}$ for Seyfert 2. The four templates of AGN dust torus are plotted in the top right panel of Figure~\ref{fig:templates} with yellow curves. The larger the column density, the higher is the nuclear contribution to the IR emission. Although the X-ray hardness ratio, i.e., the ratio between the number of counts in the $2-7$ keV band and the number of counts in the $0.5-2$ keV band, allows one to get a rough estimate of the $N_{\rm H}$ value (see \citealt{Marchesi16}), we chose to allow ${\rm N_{H}}$ to be a free parameter in the SED fitting.  

\begin{figure*}
\centering
\includegraphics[width=0.49\textwidth]{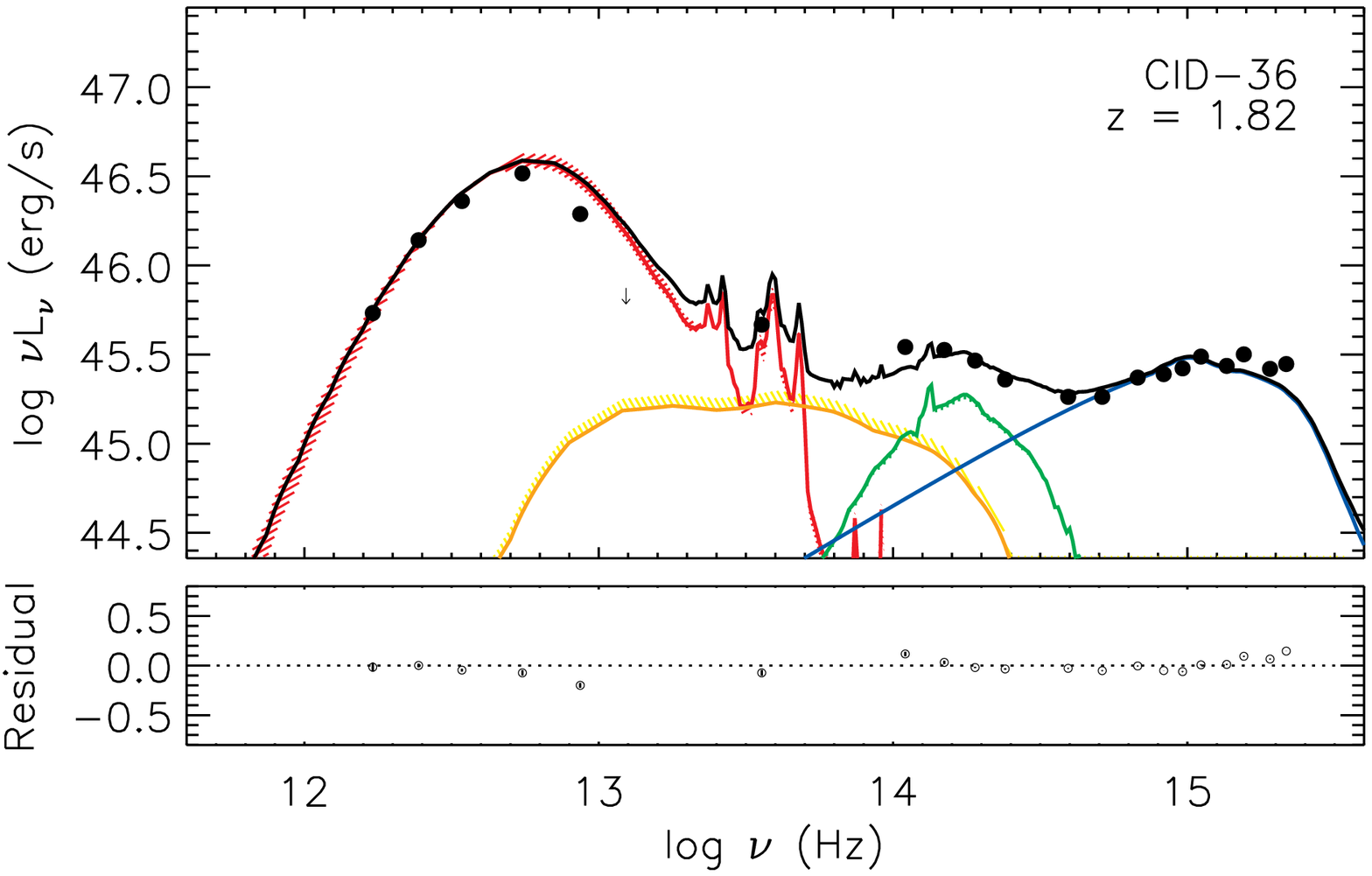}
\includegraphics[width=0.49\textwidth]{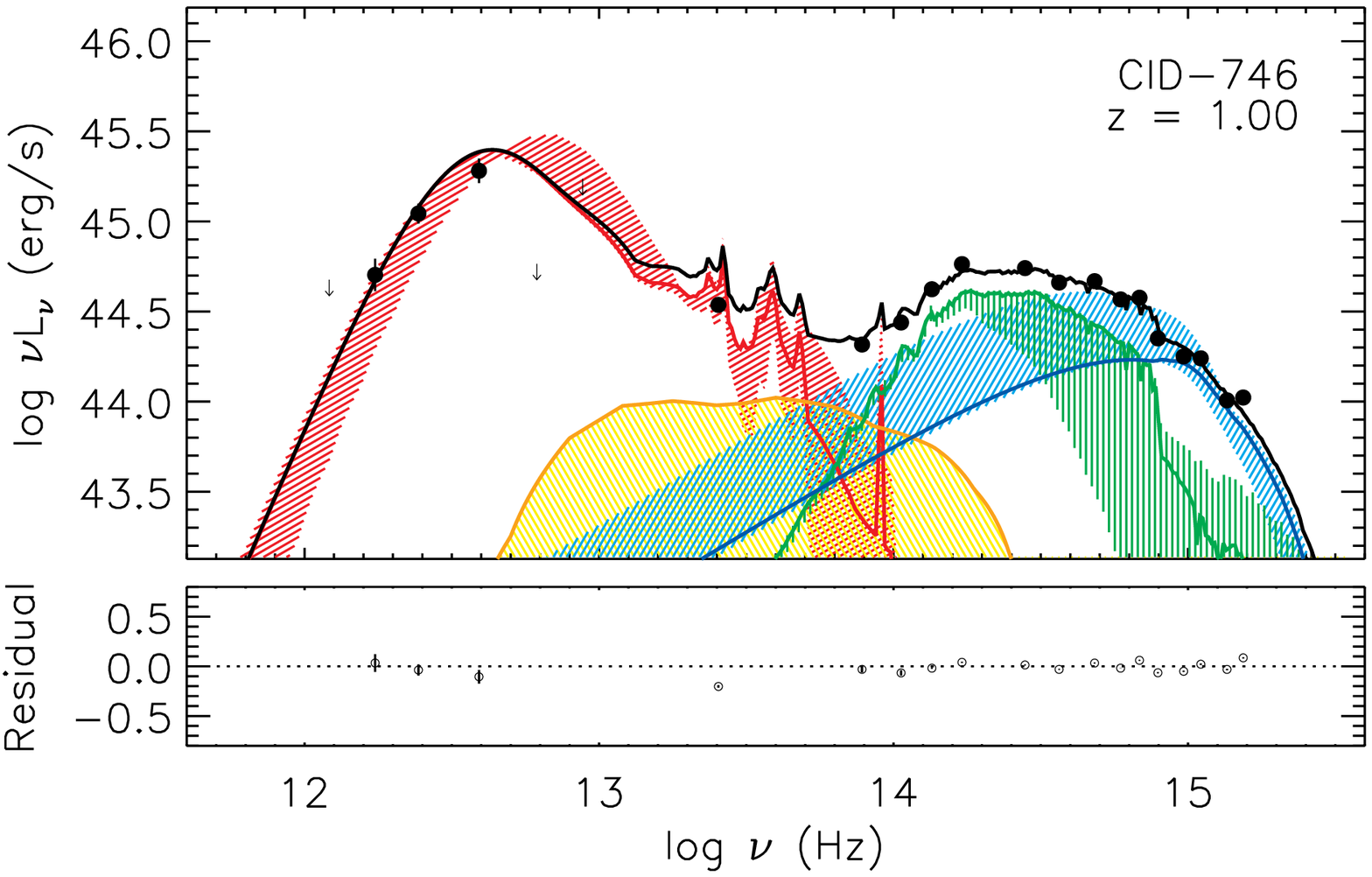}
\includegraphics[width=0.49\textwidth]{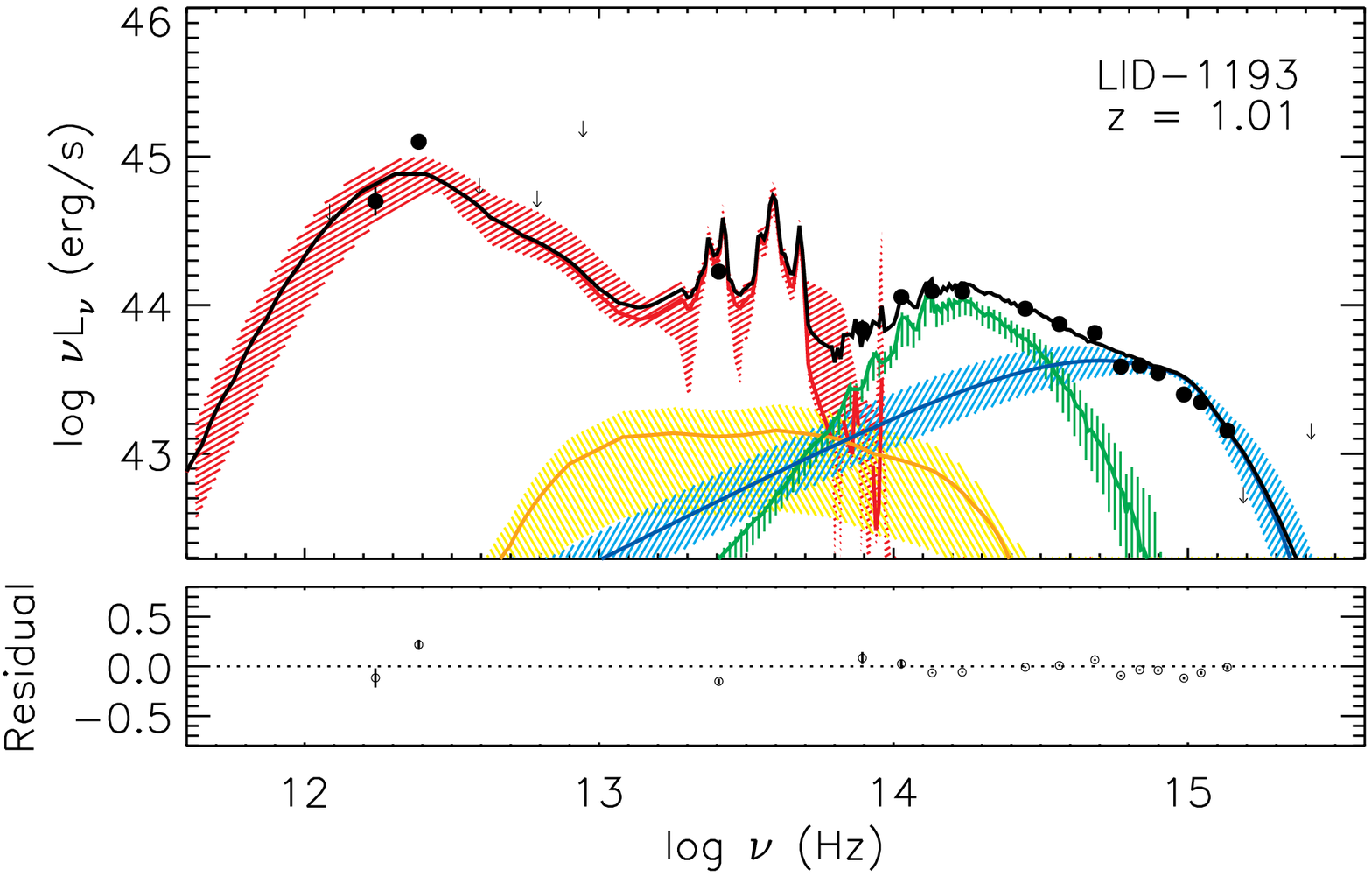}
\includegraphics[width=0.49\textwidth]{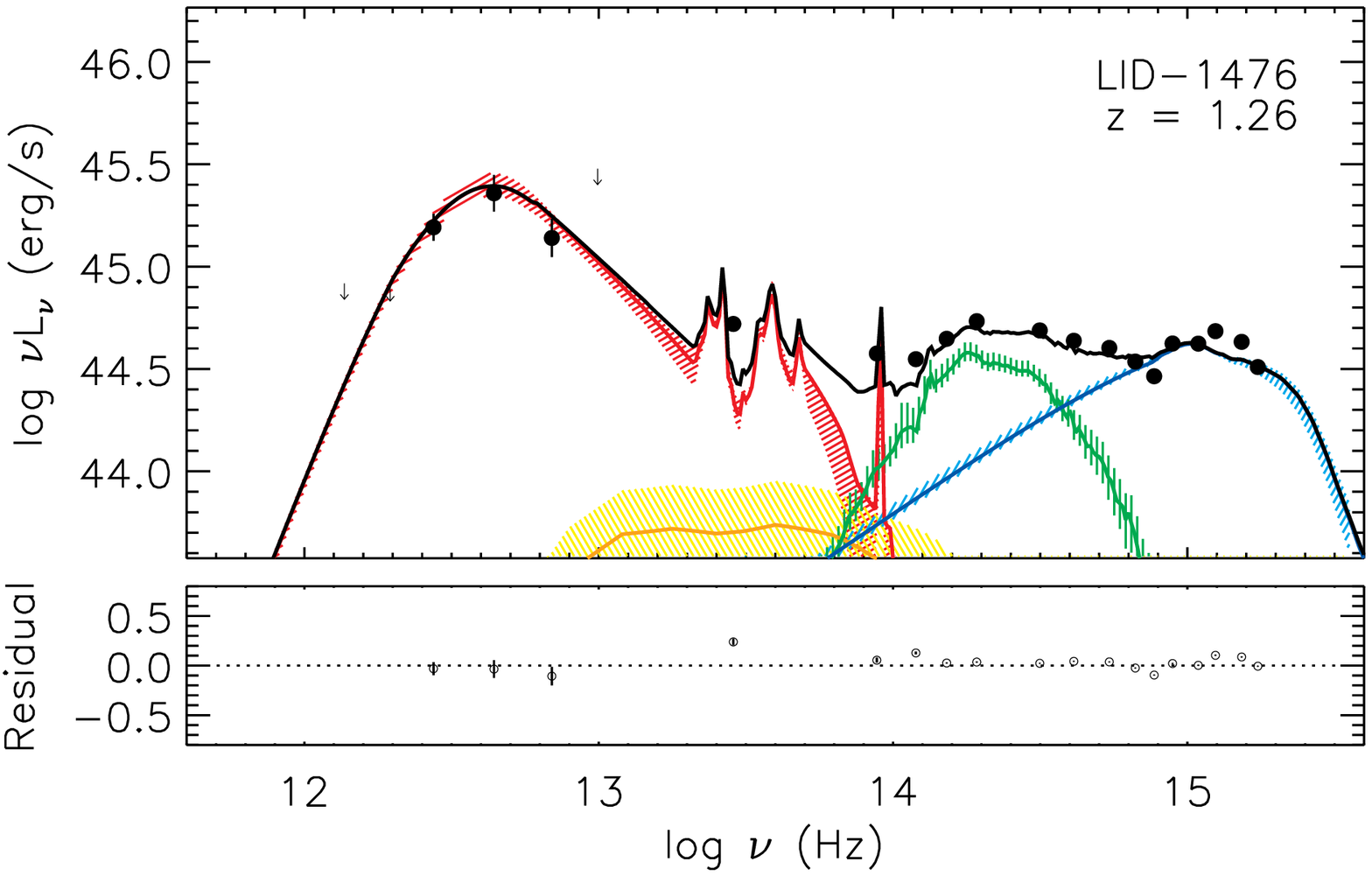}
\caption{Examples of four-component SED fits for Type 1 AGN host galaxies. The rest-frame observed photometric data (black points) and the detection limits (arrows) are shown with the best-fit model (black solid curve). The AGN BBB component (blue), galaxy template (green), AGN dust torus template (yellow), and starburst component (red) are also plotted. The residuals are shown in the lower plot of each spectrum.}
\label{fig:sed1}
\end{figure*}

A set of galaxy model templates are generated from the stellar population synthesis models of \citet{BC03} using solar metallicity and the initial mass function (IMF) of \citet{Chabrier03}. While the metallicity might have an impact on the SED fitting results, \citet{Swindle11} showed that the stellar masses are less affected by the absence of metallicity knowledge. We have built 10 exponentially decaying SFHs, where the optical SFR is defined as SFR $\propto~e^{t/\tau}$, with characteristic times ranging from $\tau=0.1$ to 30 Gyr, and a model with constant star formation. For each SFH, the SEDs are generated by models with 15 grids of ages ($t_{\rm age}$) ranging from 0.1 to 10 Gyr, with the additional constraint on each component that the age should be smaller than the age of the universe at the redshift of the source. We take into account the reddening effect using the law of \citet{Calzetti00}. We have considered $E(B-V)$ values in the range between 0 and 0.5 with steps of 0.05, and the range between 0.5 and 1 with a step of 0.1. We show some examples of stellar population templates (2640 galaxy templates in total) with various combinations of $\tau$=[0.1, 1], and t$_{\rm age}$=[0.2 Gyr, 5 Gyr] with $E(B-V)$=[0.0, 0.3, 0.5] in the bottom left panel of Figure~\ref{fig:templates} (green curves).  

In the FIR wavelength ranges, we adopted 169 starburst templates (105 from \citealt{Chary01} and 64 from \citealt{Dale02}) for fitting the cold dust emission. It has been shown that measuring the FIR luminosity from fitting the FIR region to libraries of SEDs \citep{Chary01, Dale02} gives roughly the same results as the model of a modified blackbody plus power law \citep{Casey12, U12, Lee13}. The templates of \citet{Chary01} are generated based on the SEDs of four prototypical starburst galaxies (Arp 220, ULIRG; NGC 6090, LIRG; M82, starburst; and M51, normal star-forming galaxy). The templates of \citet{Dale02} are based on 69 normal star-forming galaxies, representing a wide range of SED shapes and IR luminosities, complementing each other. A small subset of starburst templates are shown in the bottom right panel of Figure~\ref{fig:templates} as red curves. 

\begin{figure*}
\centering
\includegraphics[width=0.49\textwidth]{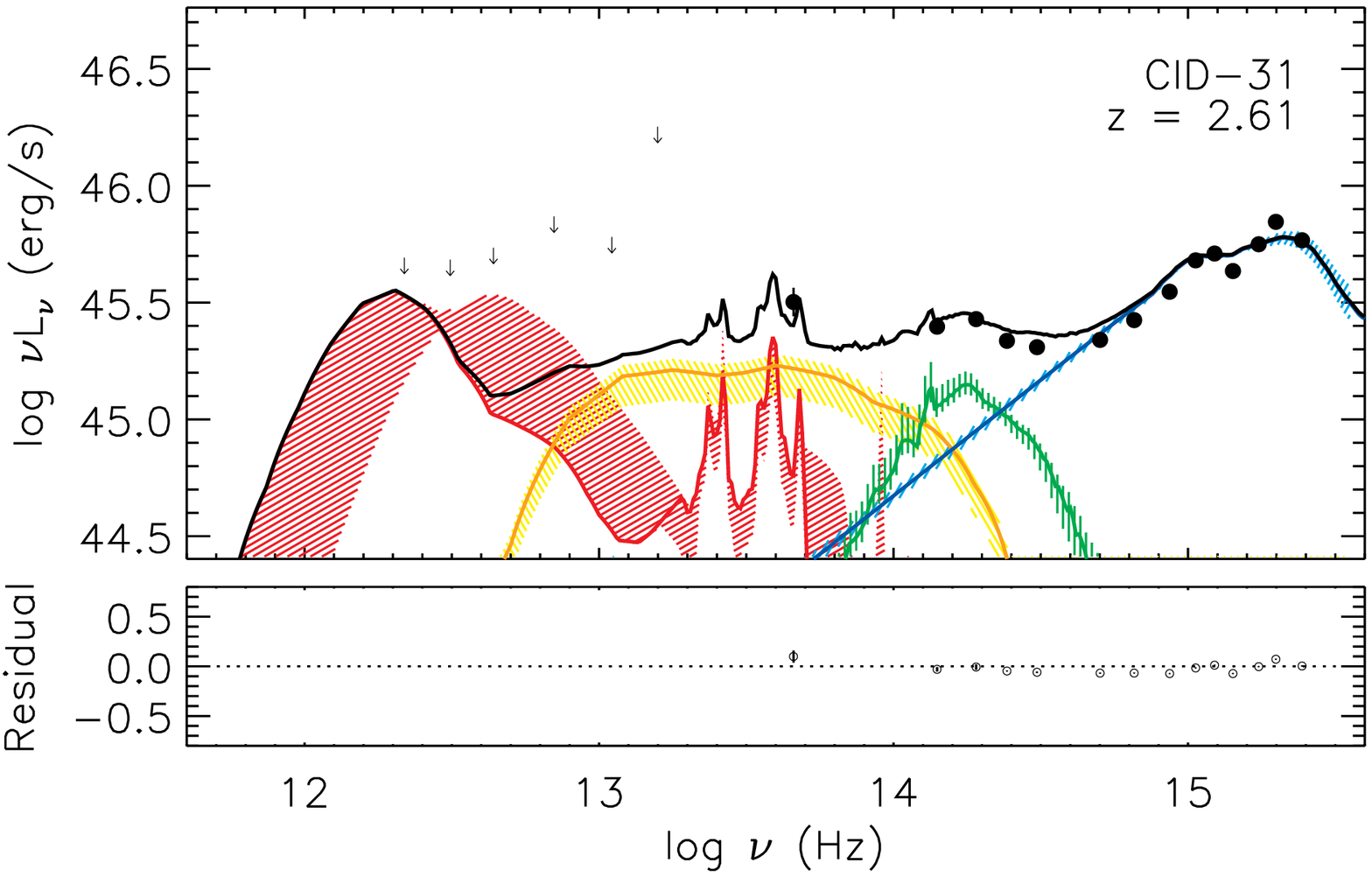}
\includegraphics[width=0.49\textwidth]{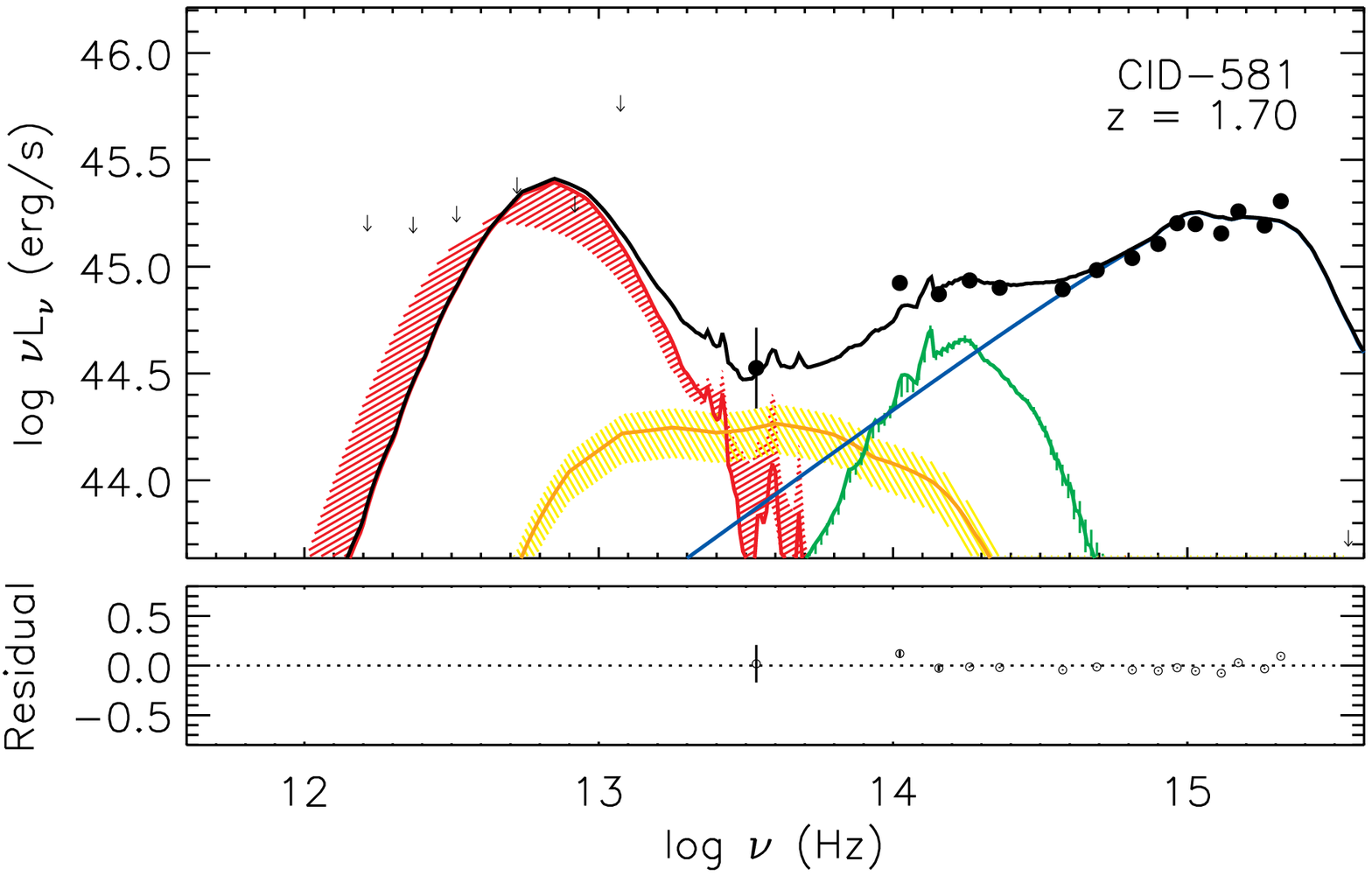}
\includegraphics[width=0.49\textwidth]{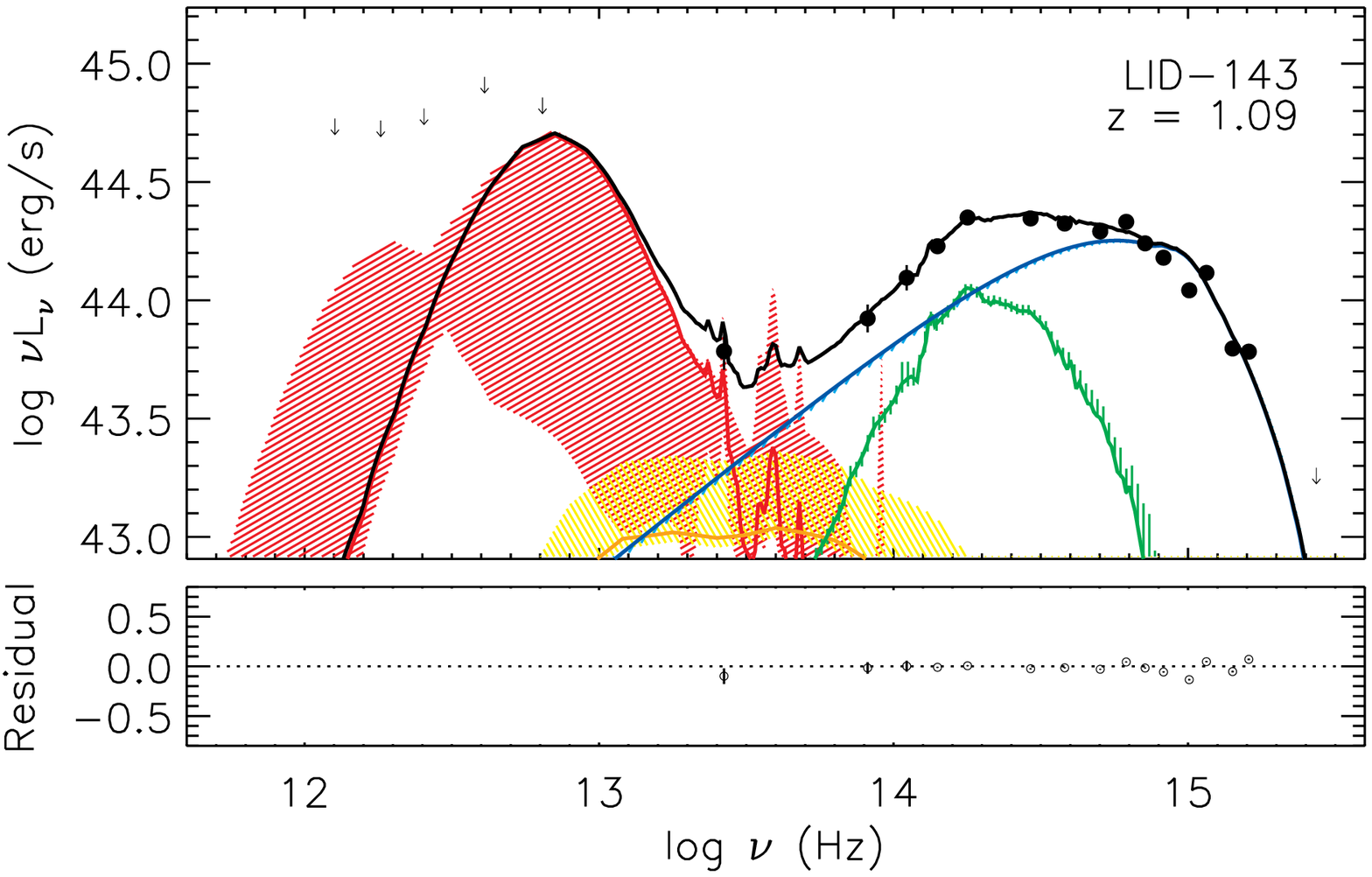}
\includegraphics[width=0.49\textwidth]{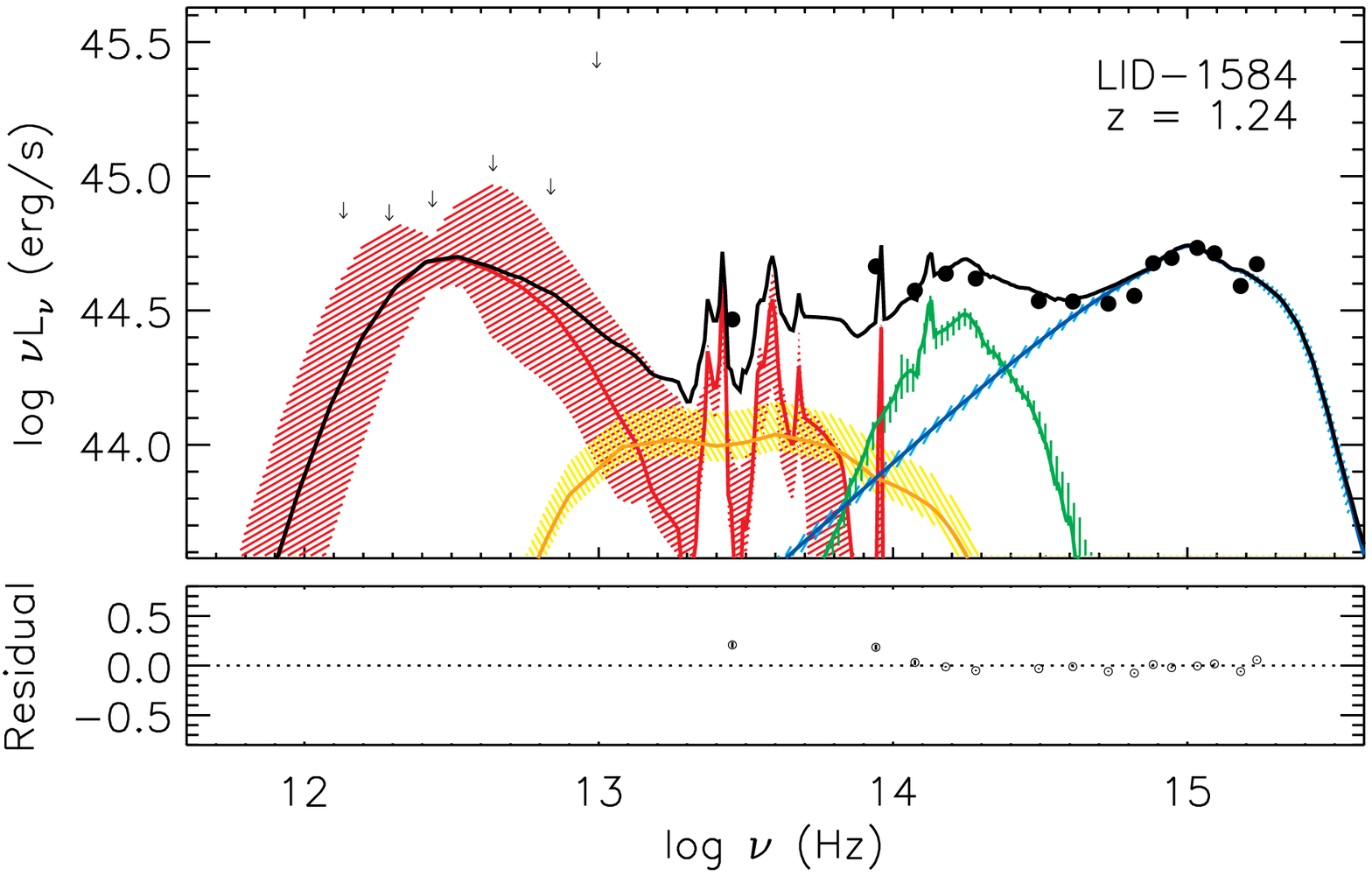}
\caption{Examples of four-component SED fits for the {\it Herschel}-undetected Type 1 AGN host galaxies. The model curves are same as in Figure~\ref{fig:sed1}.}
\label{fig:sed1n}
\end{figure*}

\subsection{Multi-component SED Fitting Procedure}\label{sec:sedfitting}

Following a similar approach to the one employed in AGNfitter (\citealt{Calistro16}; see also \citealt{Lusso13}), we develop our custom four-component SED fitting code that allows us to disentangle the nuclear emission from the stellar light, which is crucial for estimating reliable physical properties of host galaxies such as galaxy mass and SFR. We fit the observed photometric data at a fixed redshift of the source with a large grid of models obtained by combining the four-component templates described above. The observed flux can be expressed as the sum of four components as
\begin{equation*}
f_{obs} =  C_{1}f_{\rm galaxy} + C_{2}f_{\rm BBB} + C_{3}f_{\rm torus} + C_{4}f_{\rm starburst}
\end{equation*}
where the $C_{1}$, $C_{2}$, $C_{3}$, and $C_{4}$ are coefficients that reproduce the observed data by $\chi^{2}$ minimization. We set $C_{2}=0$ when fitting the SED of Type 2 AGN host galaxies (see \citealt{Suh17}). For Type 1 AGN host galaxies, we assume a non-negligible contribution from the AGN BBB component ($C_{2}\neq0$), while there could be a negligible contribution from other components. When sources have no detection in any {\it Herschel} FIR band, the {\it Herschel} survey detection limit is used to estimate the possible maximum star-forming components ($C_{4}$). Specifically, we consider the {\it Herschel} detection limits in each {\it Herschel} band (${\rm flux_{limit}}$) to make mock data points in the FIR wavelength range, assuming the flux to be ${\rm flux_{limit}}$/2 with an uncertainty $\pm~{\rm flux_{limit}}$/2, to fit the possible star-forming component (see, e.g., \citealt{Calistro16, Suh17}).

\begin{figure*}
\centering
\includegraphics[width=0.49\textwidth]{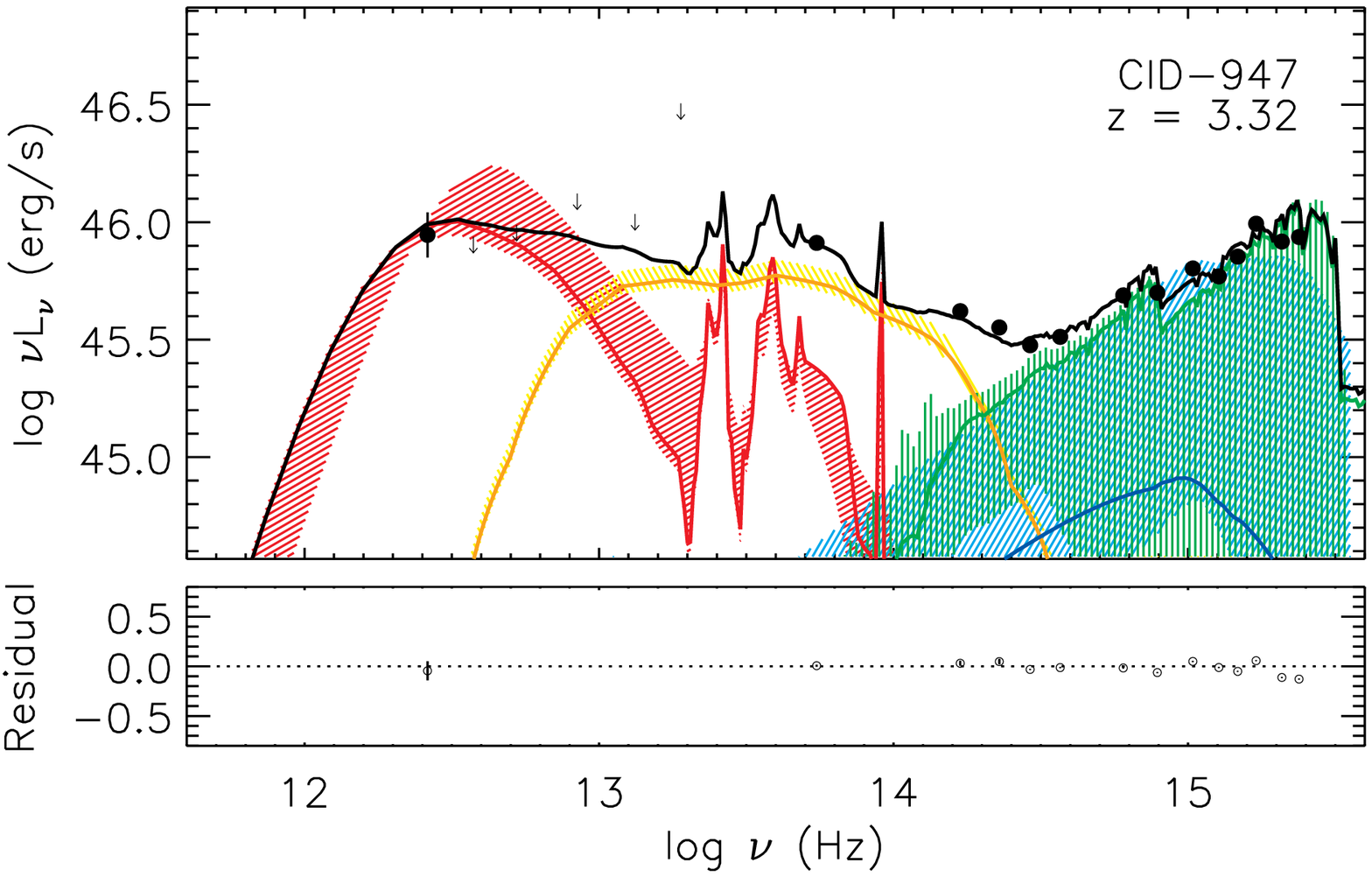}
\includegraphics[width=0.49\textwidth]{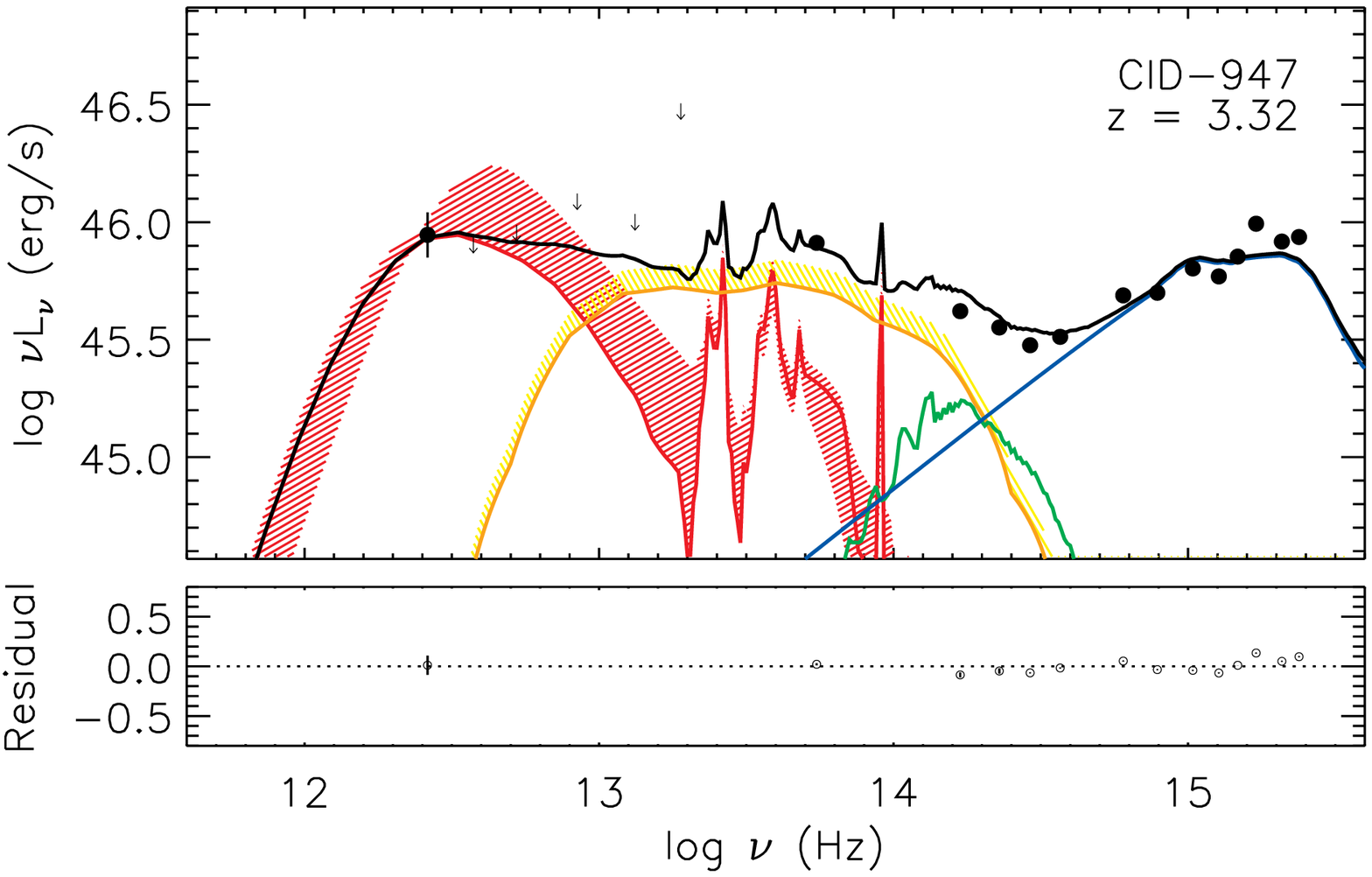}
\caption{Left: an example of SED fits for Type 1 AGN host galaxy ``CID-947'' with large uncertainties in the UV-optical wavelength range. Right: a SED fit with the constraint $C_{2} \ge C_{1}$, which makes the SED dominated by the nuclear AGN emission with a negligible galaxy contribution in the UV wavelength range. The model curves are same as in Figure~\ref{fig:sed1}.}
\label{fig:sed2}
\end{figure*}

We show examples of SED fits for Type 1 AGN host galaxies in Figures~\ref{fig:sed1}~({\it Herschel}-detected)~and~\ref{fig:sed1n}~({\it Herschel}-undetected). Examples of SED fits for Type 2 AGN host galaxies are shown in Figure 3 from \citet{Suh17}. The $\chi^{2}$ minimization is used to determine the best fit among all the possible template combinations. The rest-frame photometric data (black points) and the detection limits (arrows) are shown with the best-fit model (black solid curve). The AGN BBB template (blue), the galaxy template (green), the AGN dust torus template (yellow), and the starburst component (red) are also plotted. The residuals are also shown in the lower panel of each SED fit. In Figure~\ref{fig:sed1n}, we show examples of the SED fits for the sources that are undetected in the FIR photometry. For those {\it Herschel}-undetected sources, we show the best-fit models in the IR wavelength range with a possible star-forming component using {\it Herschel} upper limits.  

\begin{figure*}
\center
\includegraphics[width=0.49\textwidth]{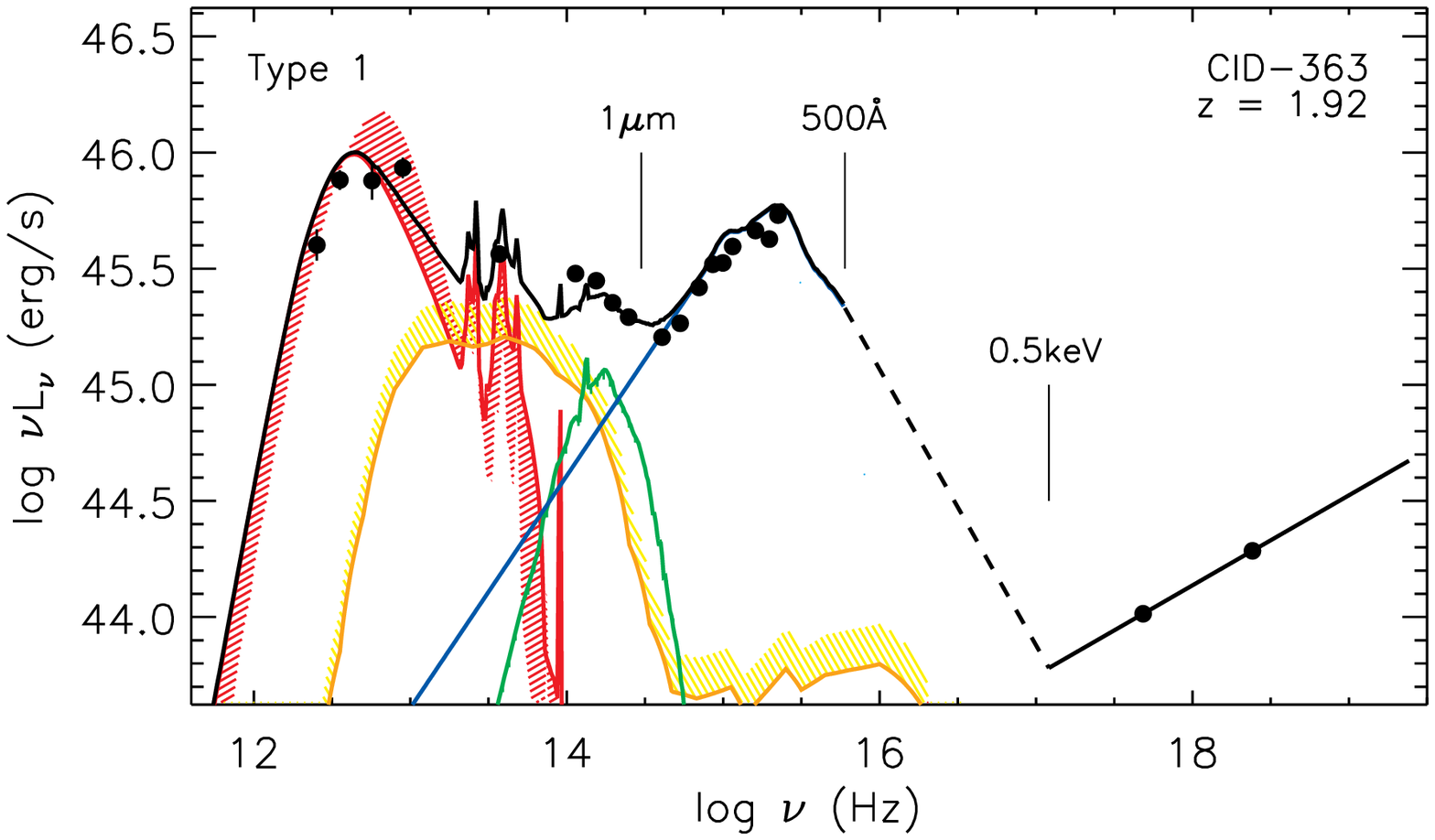}
\includegraphics[width=0.49\textwidth]{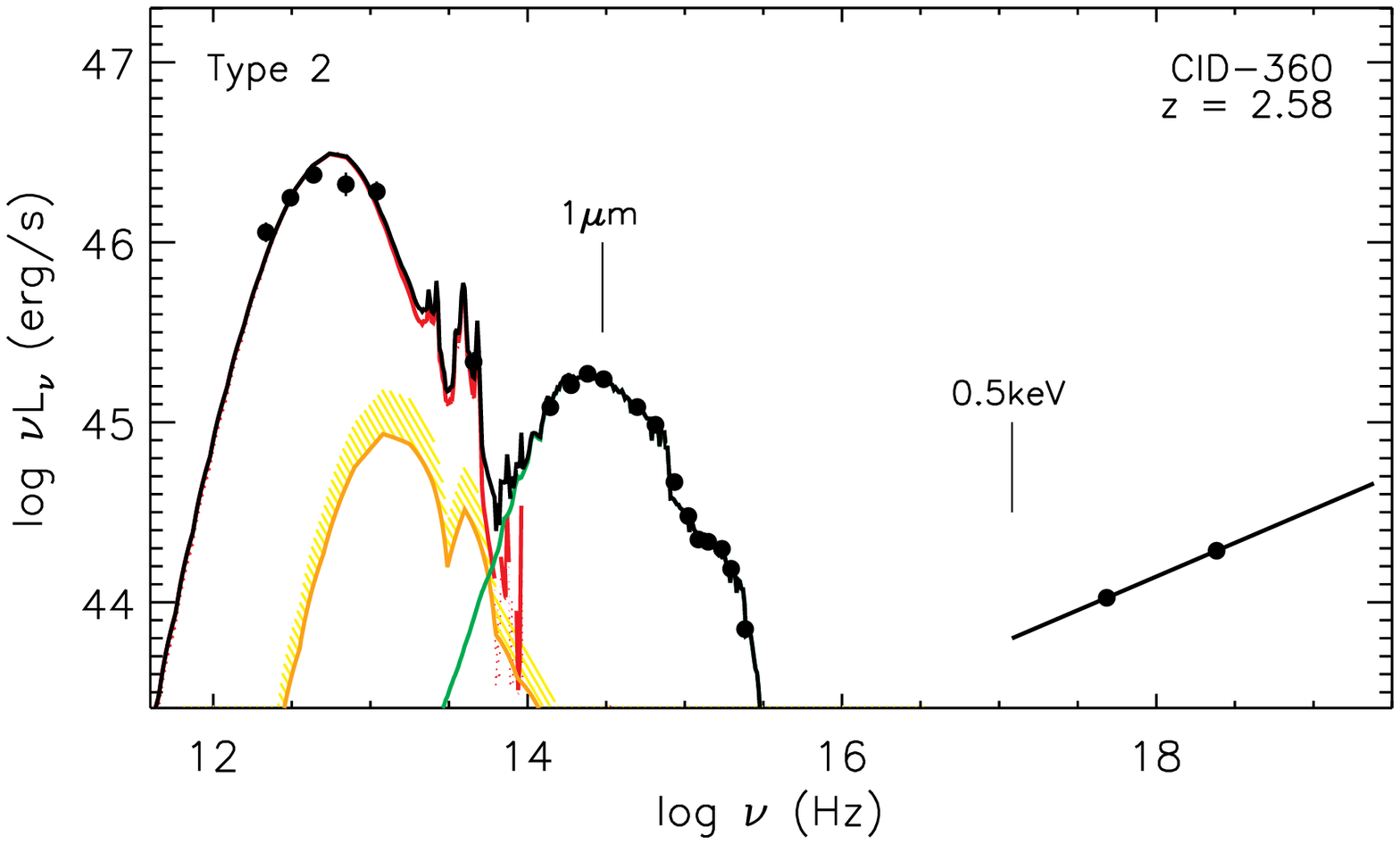}
\caption{Examples of full Type 1 (left) and Type 2 (right) AGN SEDs from FIR to X-rays. The model curves are the same as in Figure~\ref{fig:sed1}. The best-fit BBB template at 500\AA~is linearly connected to the X-ray luminosity at 0.5~keV (dashed line in the left panel). }
\label{fig:fsed}
\end{figure*}

While it is clear that both AGN BBB and galaxy components could substantially contribute in the UV-optical wavelength range, the majority of Type 1 AGN host galaxies are best fitted with old stellar populations in the IRAC bands, and the UV emission is mainly coming from the AGN BBB component with a negligible contribution from the young stellar populations, as shown in Figure~\ref{fig:sed1}~and~\ref{fig:sed1n}. However, $\sim$10\% of Type 1 AGN host galaxies show large uncertainties in the estimates of $C_{1}$ and $C_{2}$, introducing a degeneracy in the SED fitting. This implies that the fitting can produce many different probable solutions with similar $\chi^{2}$, i.e., one is a prominent AGN BBB dominating in the UV-optical ranges without contribution from the galaxy component, and the another is a negligible AGN contribution with the dominant galaxy UV emission from young stellar populations. Since there is an observed correlation between the X-ray and the UV-optical emission for AGNs (i.e., $\alpha_{\rm ox}$; see, e.g., \citealt{Tananbaum79, Lusso10, Lusso16}), the UV-optical emission should be come from the AGN contribution within the intrinsic scatter. Thus, as a further constraint on these cases we enforce $C_{2} \ge C_{1}$, to ensure that the AGN BBB component dominates in the UV bands with only a small contribution from the host galaxy's young stellar populations.  We show the example case where both AGN BBB and galaxy components could dominates in the UV-optical wavelength range with large uncertainties in the left panel of Figure~\ref{fig:sed2}. With the constraint $C_{2} \ge C_{1}$, the AGN emission dominates the galaxy's lights in the UV-optical wavelength range (right panel in Figure~\ref{fig:sed2}). We confirm that the monochromatic luminosity at rest-frame 2500\AA~of the best-fitting AGN BBB component correlates with the X-ray luminosity within the scatter of $\sim0.4$ dex.

\section{SED fitting results}

\subsection{AGN Luminosities}\label{sec:AGN_lum}

We compute the relevant nuclear luminosities from the different components of the SED by integrating the best-fit model over a specific range. Specifically, we compute the absorption-corrected intrinsic total X-ray luminosity ($\rm L_{0.5-100~keV}$) by integrating over the range $0.5-100$ keV, assuming a photon index $\Gamma=1.8$. We estimate the intrinsic nuclear emission in the UV-optical range by integrating the best-fit AGN BBB template (${\rm L_{BBB}}$) over the range ${\rm 500\AA-1\mu m}$, taking into account the AGN reddening. The AGN torus luminosity (${\rm L_{torus}}$) is obtained by integrating the dust torus template from 1 to 1000 $\mu$m. 

The total AGN luminosity for Type 1 AGNs is computed as the sum of the $\rm L_{0.5-100~keV}$ and ${\rm L_{BBB}}$ using an approach similar to that of \citet{Lusso13}. We linearly connect the AGN BBB luminosity at 500${\rm \AA}$ to the luminosity corresponding to the absorption-corrected X-ray spectrum at 0.5 keV. The resulting total luminosity for Type 1 AGNs is integrated from 1 $\mu$m to 100 keV. For Type 2 AGNs, the total AGN luminosity is computed as the sum of the $\rm L_{0.5-100~keV}$ and ${\rm L_{torus}}$ using an approach similar to that of \citet{Lusso11} (see also \citealt{Pozzi07}). To convert the IR luminosity into a proxy for the intrinsic nuclear luminosity, we consider the geometry of the torus and its orientation by applying the following correction factors (see \citealt{Pozzi07, Lusso11}): the first correction is related to the covering factor, which represents the fraction of the primary UV-optical radiation intercepted by the torus ($\sim$1.5; see, e.g., \citealt{Gilli07}), and the second correction is due to the anisotropy of the IR dust emission, which is a function of the viewing angle ($\sim$1.3; see, e.g., \citealt{Vasudevan10, Lusso11}). Examples of the full SED for Type 1 (left) and Type 2 (right) AGNs from FIR to X-rays are shown in Figure~\ref{fig:fsed}. The dashed line in the left panel represents the extrapolation between the AGN BBB luminosity at 500\AA~and the X-ray luminosity at 0.5 keV. 

Given that the total luminosity for Type 1 and Type 2 AGNs is computed by integrating different ranges of the SED, we compute the bolometric luminosity using the X-ray luminosity for both samples for consistency. The bolometric luminosity (${\rm L^{X-ray} _{bol}}$) is derived from the absorption-corrected rest-frame $2-10$ keV luminosity (see \citealt{Marchesi16}) with the luminosity-dependent bolometric correction factor described in \citet{Marconi04}. For sources that are not detected in the hard band but are detected in the full band, ${\rm L_{2-10~keV}}$ values are estimated using upper limits. 2826 sources have been detected in the full band (2423 and 2264 in the soft and hard bands, respectively). \citet{Marconi04} derived the bolometric corrections from an AGN template spectrum of optical, UV, and X-ray luminosities radiated by the accretion disk and hot corona (see also \citealt{Vignali03, Hopkins07, Lusso12}). They considered only the AGN-accretion-powered luminosity, neglecting the luminosity reprocessed by the dust in IR luminosities. Despite some difference between the luminosity-dependent bolometric correction factors among different studies (e.g., \citealt{Hopkins07, Lusso12}), the same trend of increasing bolometric correction at increasing bolometric luminosity is observed within the scatter. The scatter is $\sim0.1$ for X-ray luminosities.

\begin{figure}
\center
\includegraphics[width=0.5\textwidth]{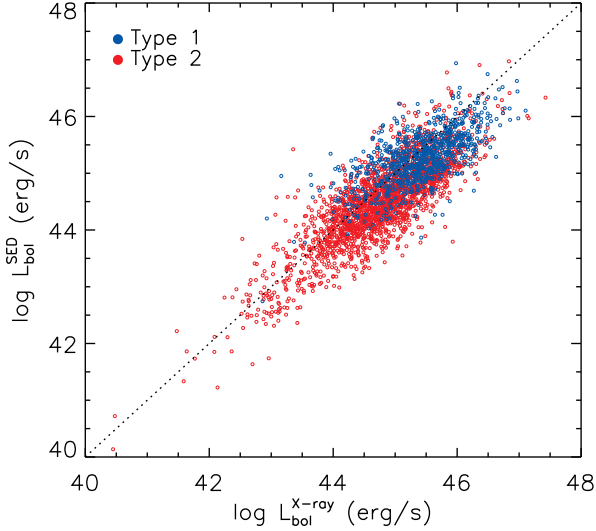}
\caption{The total AGN luminosity derived from the SED fitting versus the AGN bolometric luminosity derived from X-ray luminosity for Type 1 (blue) and Type 2 (red) AGNs. The dotted line denotes a one-to-one relation.}
\label{fig:Lbol}
\end{figure}

In Figure~\ref{fig:Lbol}, we show the total AGN luminosity derived from the SED fitting with respect to ${\rm L^{X-ray} _{bol}}$, where the one-to-one correlation is plotted as a dotted line for reference. Type 1 and Type 2 AGNs are indicated with blue and red symbols, respectively. The total AGN luminosities derived from the SED fitting are on average slightly lower than ${\rm L^{X-ray} _{bol}}$, with median offsets of -0.1 dex and -0.2 dex for Type 1 and Type 2 AGNs. We find a 1$\sigma$ dispersion of $\sim$0.4 dex for both Type 1 and Type 2 AGNs. We note that $\sim$3\% of the sample show an AGN luminosity of ${\rm L^{X-ray} _{bol}~erg~s^{-1}<43}$, for which the observed X-ray emission could be contaminated by X-ray binaries (XRBs) and/or hot interstellar medium (ISM) gas. With the X-ray hardness ratio (HR) as a proxy for which type of X-ray emission we are observing, we confirm that all the sources have an HR (i.e., HR=(H-S)/(H+S), where H and S are the count rates in the 0.5--2 keV and 2--7 keV X-ray bands) above that of thermal emission (i.e., typically HR$\lesssim-0.8$ with photon index $\Gamma=3$) but consistent with hard X-ray emission, which supports the AGN nature (see also \citealt{Mezcua18}). 

\subsection{Stellar Mass}\label{sec:stellar_mass}

\begin{figure*}
\center
\includegraphics[width=0.9\textwidth]{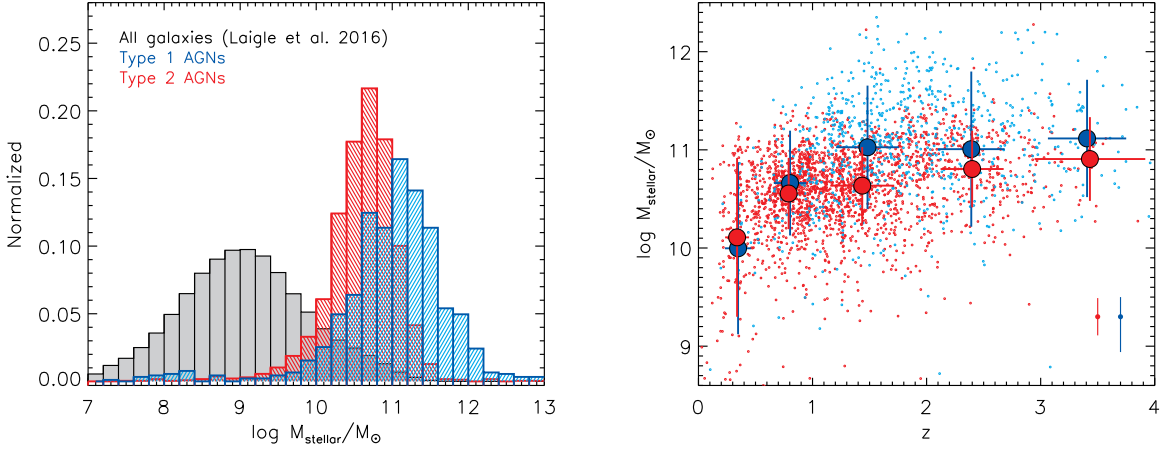}
\caption{Left: stellar mass histogram of our sample of Type 1 (blue) and Type 2 (red) AGN host galaxies, normalized to the total area. The distribution of all galaxies from the COSMOS catalog \citep{Laigle16} is shown by the gray histogram for comparison. Right: stellar mass versus redshift distribution. The individual sources are indicated with small symbols. Large symbols indicate mean values of stellar mass for Type 1 (blue) and Type 2 (red) AGN host galaxies, in different redshift bins. We also show the typical uncertainties in the bottom right corner.}
\label{fig:Ms}
\end{figure*}

\begin{figure}
\center
\includegraphics[width=0.5\textwidth]{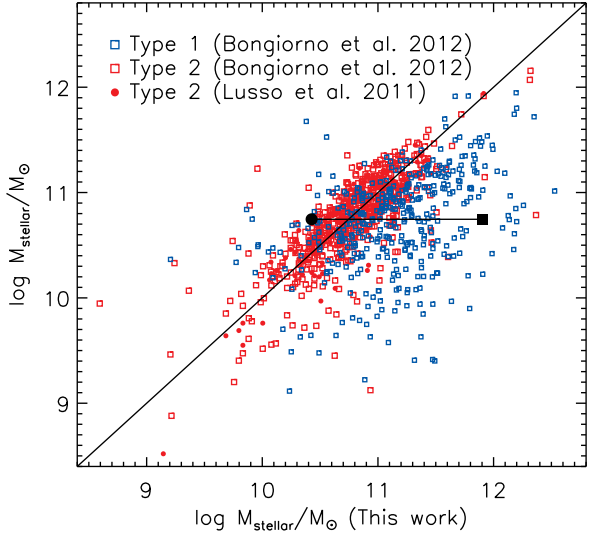}
\caption{Comparison between stellar masses derived from our SED fitting and those from \citet[filled red circles]{Lusso11} and \citet[empty squares]{Bongiorno12}. The black line denotes a one-to-one relation. We show the stellar mass for Type 1 AGN ``CID-947" derived from two different SED fits with a black circle (from the left panel of Figure~\ref{fig:sed2}) and a black square (from the right panel of Figure~\ref{fig:sed2}).}
\label{fig:Ms_comp}
\end{figure}

We derive a probability distribution function (PDF) for the host stellar mass with the likelihood, exp($-0.5\chi^{2}$), considering any possible combination of SED parameters, which includes the age since the onset of star formation, the $e$-folding time $\tau$ for exponential SFH models, and the dust reddening. A detailed description is presented in \citet{Suh17}. Figure~\ref{fig:Ms} shows the stellar mass distributions for our sample of AGN host galaxies. In the left panel, the normalized distributions of stellar masses for Type 1 and Type 2 AGN host galaxies are indicated by blue and red histograms, respectively. For comparison, the stellar mass distribution of all the galaxies in the COSMOS field \citep{Laigle16} is shown by the gray histogram. In the right panel of Figure~\ref{fig:Ms}, we show the redshift evolution of stellar masses for our sample of Type 1 (blue) and Type 2 (red) AGN host galaxies. Individual sources are indicated with small symbols. Large symbols represent the mean and the standard deviation. The typical uncertainties for the stellar masses for Type 1 (blue) and Type 2 (red) AGNs are shown in the bottom right corner. 

The stellar mass of our sample ranges from $\sim10^{9}$ to $10^{13}$ M$_{\odot}$, peaking at higher masses ($\sim5\times10^{10}$ M$_{\odot}$) than normal galaxies ($\sim10^{9}$ M$_{\odot}$), consistent with previous studies (e.g., \citealt{Lusso11, Bongiorno12}). The stellar mass distribution of Type 1 AGN host galaxies peaks at ${\rm log~M_{stellar}/M_{\odot}\sim11.07}$ with a dispersion of 0.73 dex, while those of Type 2 AGN host galaxies peak at lower masses of ${\rm log~M_{stellar}/M_{\odot}\sim10.64}$ with a dispersion of 0.51 dex. However, we should note here that Type 1 AGN host galaxies could be biased toward old stellar populations due to the degeneracy in the SED fitting, which gives rise to a bias toward higher stellar masses (see Section~\ref{sec:sedfitting}). While \citet{Bongiorno12} suggested that there is no significant difference between the mass distributions of Type 1 and Type 2 AGN host galaxies, the measurement of the stellar mass for Type 1 AGN host galaxies has considerably large uncertainties. 

In Figure~\ref{fig:Ms_comp}, we show the comparison of the stellar masses obtained from our SED fitting with the results from \citet[Type 2, filled red circles]{Lusso11} and \citet[Type 1 (blue) and Type 2 (red), empty squares]{Bongiorno12} based on their SED fitting on the $XMM$-COSMOS data set. We find good agreements on the stellar masses for Type 2 AGN host galaxies with the 1$\sigma$ dispersions of 0.27 dex \citep{Lusso11} and 0.3 dex \citep{Bongiorno12}, respectively (see \citealt{Suh17}). However, there is a large disagreement on the stellar mass for Type 1 AGN host galaxies (blue symbols). As explained in Section~\ref{sec:sedfitting}, we enforce the domination of the AGN emission over the host galaxy light in the UV bands, so our stellar mass measurement for Type 1 AGN host galaxies could be biased toward an upper limit on the stellar mass. We show an example case of how uncertain the stellar mass of Type 1 AGN host galaxy ``CID-947" derived from two different SED fits in Figure~\ref{fig:Ms_comp} (black symbols). The black circle indicates the stellar mass derived from the SED fit in the left panel of Figure~\ref{fig:sed2}, for which the UV emission is dominated by the galaxy's young stellar populations (green solid curve), i.e., a lower limit on the stellar mass. The black square indicates the stellar mass we adopted in this study, derived from the SED fit in the right panel of Figure~\ref{fig:sed2}, for which the AGN emission dominates in the UV-optical wavelength range. 

\subsection{Star Formation Rate}

The SFR is estimated using the total star-forming IR luminosity by integrating the best-fit starburst template. While combining the contributions from UV and IR luminosity provides an estimate of both the obscured and unobscured SFRs (see, e.g., \citealt{Bell05, Arnouts13, Lee15, Suh17}), we use only IR luminosities because the accretion disk emission contributes strongly in the UV range, introducing a degeneracy between the UV emission from star formation and that from the central AGN. \citet{Lee15} found that for sources with SFR${\rm >50M_{\odot}~yr^{-1}}$, the IR contribution dominates the total SFR, contributing as much as $\sim90\%$ of it. We thus derived the total SFRs by using the relation from \citet{Kennicutt88} for a \citet{Chabrier03} IMF. 

\vspace{0.1in} \hspace{0.1in} ${\rm SFR_{total}~(M_{\odot}/yr)} = 10^{-11} \times L^{\rm SF} _{\rm IR}/L_{\odot}$ 

\vspace{0.1in}
where $L^{\rm SF} _{\rm IR}$ is the total rest-frame IR luminosity, which is integrated between 8 and 1000 $\mu$m from the starburst template. Since a significant fraction ($\sim$73\%) of our sample are faint in the FIR photometry, to account for the {\it Herschel}-undetected sources, we derive upper limits on their SFRs by assuming a possible star-forming IR luminosity from the best-fit starburst template using {\it Herschel} detection limits (see Section \ref{sec:sedfitting}). For Type 2 AGN host galaxies, \citet{Suh17} derived the lower limits on SFRs using only UV luminosity from the best-fit galaxy template for the {\it Herschel}-undetected sources, and found that the average difference between the upper and lower limits on SFRs is $\sim0.3$ dex. 

\begin{figure}
\center
\includegraphics[width=0.5\textwidth]{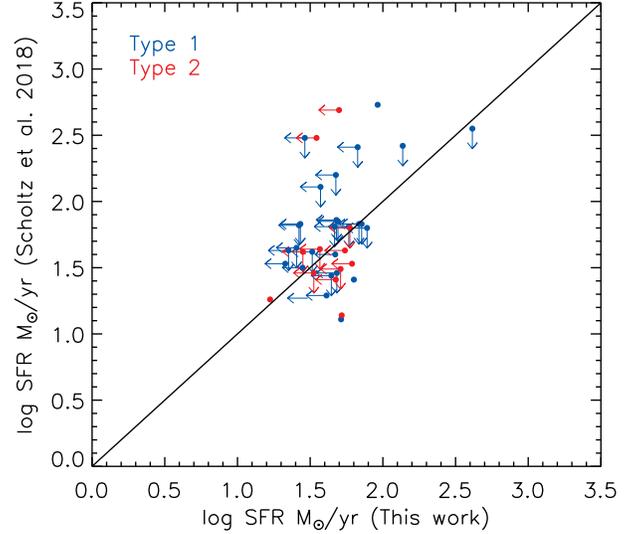}
\caption{Comparison of SFRs derived from our SED fitting with those from \citet{Scholtz18}. The downward and left-pointing arrows indicate the upper limit on SFRs of \citet{Scholtz18} and this work (i.e., {\it Herschel}-undetected sources), respectively. The black line denotes a one-to-one relation. }
\label{fig:SFR}
\end{figure}

Recently, \citet{Scholtz18} used sensitive ALMA 870$\mu$m continuum observations in combination with data from {\it Spitzer} and {\it Herschel} to compute SFRs for X-ray-selected AGNs in CDF-S and COSMOS fields. In Figure~\ref{fig:SFR}, we show the comparison of SFRs obtained from our SED fitting with the results from \citet{Scholtz18}. The blue and red symbols indicate Type 1 and Type 2 AGN host galaxies, respectively. We show the upper limits on SFRs with downward arrows \citep{Scholtz18} and left-pointing arrows ({\it Herschel}-undetected sources). We find relatively good agreements on the SFRs with a 1$\sigma$ dispersion of $\sim$0.2 dex. 

\section{Results}

\begin{figure}
\center
\includegraphics[width=0.5\textwidth]{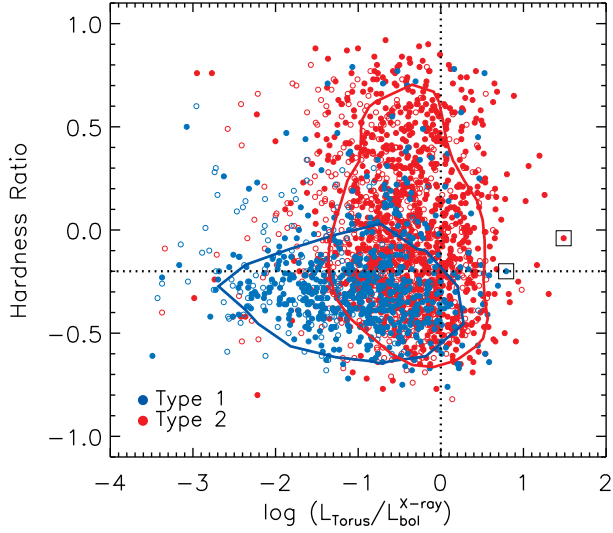}
\caption{Hardness ratio (HR=(H-S)/(H+S)) versus covering factor (CF=${\rm L_{torus}/L^{X-ray} _{bol}}$) for our sample of Type 1 (blue) and Type 2 (red) AGNs with contours at the 1$\sigma$ level. Filled symbols indicate MIPS 24$\mu$m detected sources while empty symbols represent 24$\mu$m undetected sources. The dotted horizontal line indicates HR=$-0.2$, which can be used as a threshold for X-ray absorbed objects from \citet{Hasinger08}. We show the example SEDs for the AGNs with MIR excess (black large squares) in Figure~\ref{fig:sed3}.}
\label{fig:CFHR}
\end{figure}

\begin{figure}
\centering
\includegraphics[width=0.5\textwidth]{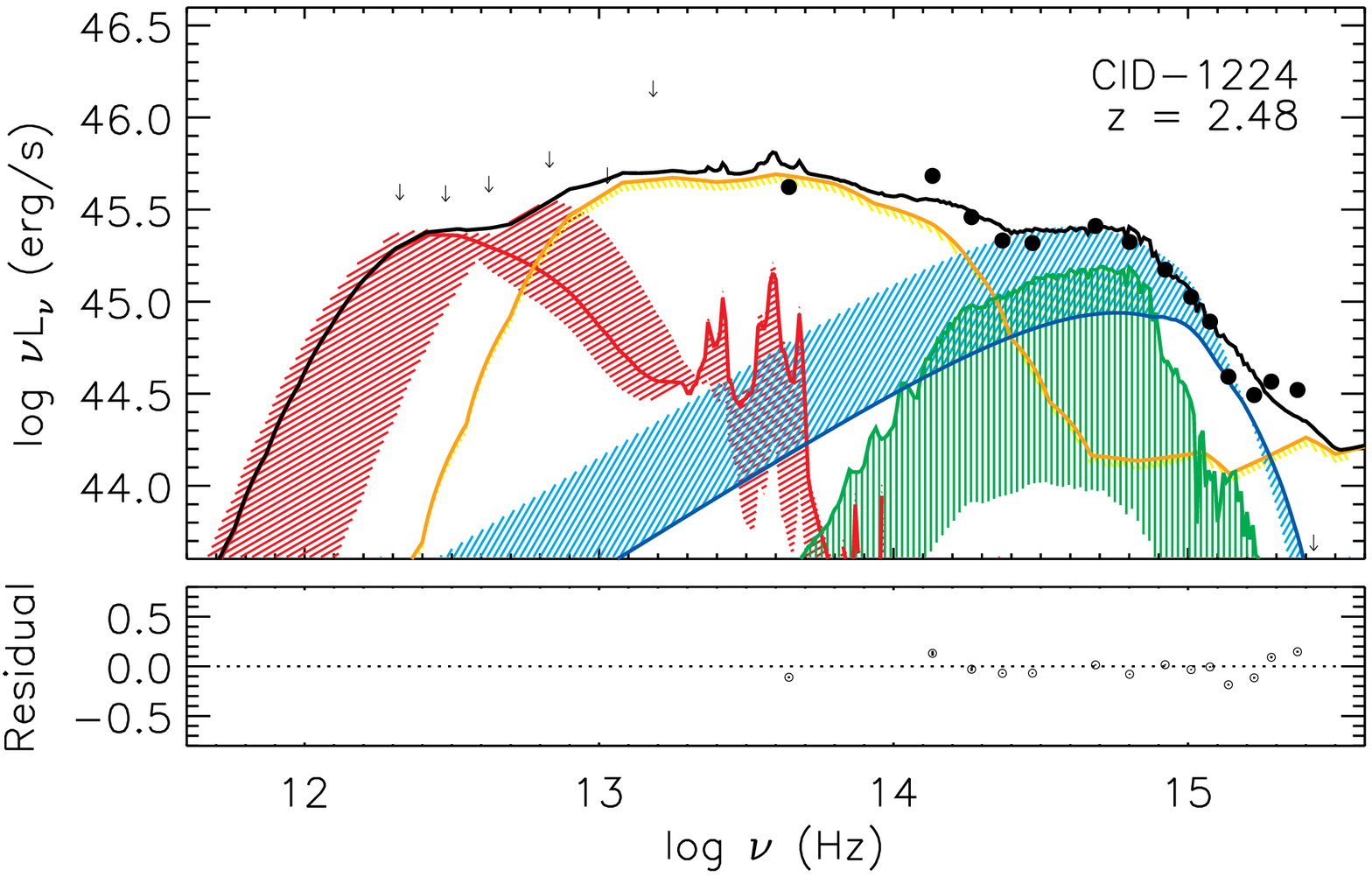}
\includegraphics[width=0.5\textwidth]{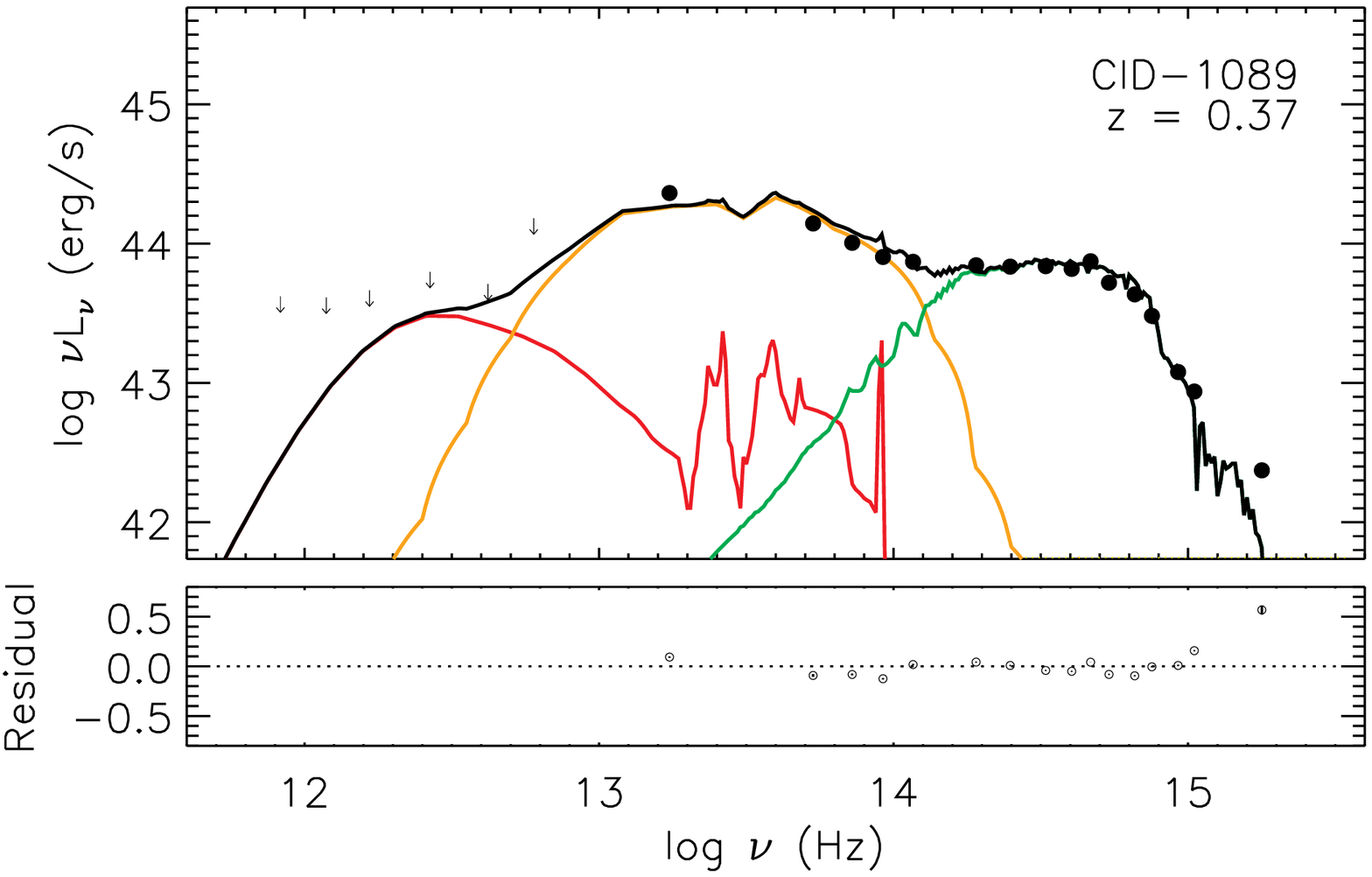}
\caption{Examples of SED fits for the Type 1 (top) and Type 2 (bottom) AGNs with MIR excess (${\rm L_{torus}/L^{X-ray} _{bol}}>1$; black squares in Figure~\ref{fig:CFHR}). The model curves are same as in Figure~\ref{fig:sed1}.}
\label{fig:sed3}
\end{figure}

\subsection{AGN Absorption and Obscuration}

\begin{figure*}
\center
\includegraphics[width=1\textwidth]{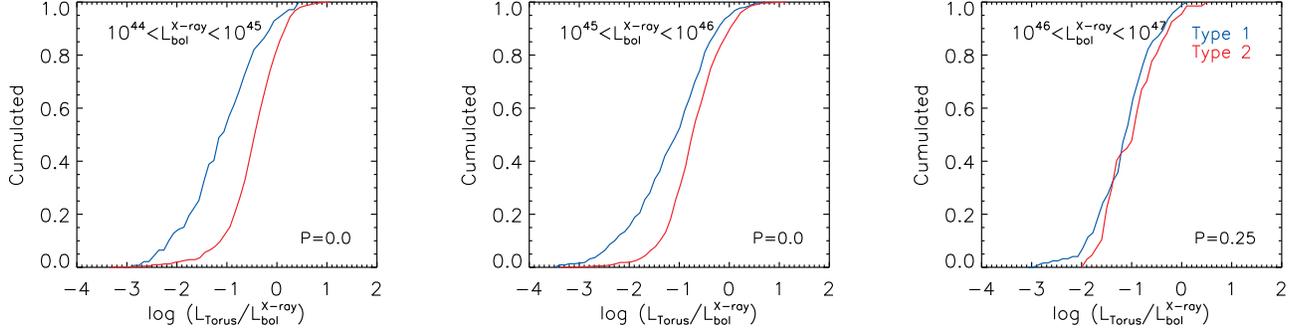}
\caption{Cumulative distributions of covering factors for Type 1 (blue) and Type 2 (red) AGNs. In order to minimize a selection bias with AGN luminosity, we show Type 1 and Type 2 AGNs in the same AGN luminosity bins. The K-S test discriminates between these two distributions at $>99.999$\% confidence level.}
\label{fig:KStest}
\end{figure*}

The obscuration of AGNs is particularly important for understanding the structure of the dust surrounding the nucleus. Having decomposed the torus emission from the SED fitting, it is of interest to study the absorption resulting from the dust and its connection to the gas absorption in the X-ray emission. We compute the dust covering factor, which is represented by the ratio of the dusty torus emission to the AGN bolometric luminosity (e.g., \citealt{Maiolino07, Treister08, Lusso13}). We show the distribution of hardness ratios HR as a function of the covering factors (i.e., CF=${\rm L_{torus}/L^{X-ray} _{bol}}$) for our sample of Type 1 (blue) and Type 2 (red) AGNs with contours at the 1$\sigma$ level in Figure~\ref{fig:CFHR}. The typical uncertainties for ${\rm L_{torus}}$ are $\sim0.10$ dex for Type 1 AGNs and $\sim0.18$ dex for Type 2, respectively. We plot the hardness ratio of $-0.2$ (dotted horizontal line), which can be used as a threshold that distinguishes between X-ray absorbed and unabsorbed objects from \citet{Hasinger08}. 

The majority of Type 1 AGNs are unobscured in X-rays and have a narrow distribution of HRs with an average of $\sim-$0.3, and an overall wide spread in CFs with the average value of $\sim$0.06. The average HR and CF values for Type 2 AGNs are $\sim-$0.09 and $\sim$0.2, respectively. We do not find a clear separation of CFs between Type 1 and Type 2 AGNs, while the CF distribution of Type 1 AGNs extends toward lower values. It is interesting to note that there are AGNs with MIR excess (${\rm L_{torus}/L^{X-ray} _{bol}}>1$), which could have underestimated ${\rm L^{X-ray} _{bol}}$ because of the weak X-ray emission due to the high absorption. These sources could potentially be heavily obscured sources (i.e., Compton-thick AGNs). We found that $\sim$5\% of Type 1 AGNs and $\sim$13\% of Type 2 AGNs are possibly heavily obscured (${\rm L_{torus}/L^{X-ray} _{bol}}>1$). We show the example SEDs for the Type 1 (top) and Type 2 (bottom) AGNs with MIR excess (black large squares in Figure~\ref{fig:CFHR}) in Figure~\ref{fig:sed3}.

Our results are consistent with the previous study of \citet{Mateos16}, which also found a very strong overlap in CF distributions between the different types of AGNs. While there is a similar classification between the X-ray absorption properties and the UV/optical spectroscopy of AGNs, it is also well known that the gas absorption in X-rays and the dust extinction are not always correlated (e.g., \citealt{Merloni14}), which is challenging to explain with the standard orientation-based unification model. Based on the X-ray spectral fitting, \citet{Marchesi16} reported that $\sim15\%$ of Type 1 AGNs in CCLS are X-ray obscured and $\sim18\%$ of Type 2 AGNs are X-ray unobscured. It has been suggested that the obscuring dust in AGNs is not uniformly distributed but is clumpy and allows emission from the broad-line region to escape from the torus without being obscured (see, e.g., \citealt{Netzer15}). This implies that the observed differences between Type 1 and Type 2 AGNs are due not only to simple orientation effects but also due to the clumpy structure of the dusty torus. 

Figure~\ref{fig:KStest} shows the cumulative distribution of covering factors for Type 1 (blue) and Type 2 (red) AGNs. In order to avoid selection bias with respect to AGN luminosity, we bin both Type 1 and Type 2 AGNs to the same luminosity ranges. Based on the Kolmogorov-Smirnov test (K-S test), we find that Type 1 and Type 2 AGNs at ${\rm L^{X-ray} _{bol}~erg~s^{-1}<46}$ are not consistent with being drawn from the same parent population with a confidence level higher than 99.999\% ($p=0.0$). However, for the high-luminosity bin (${\rm 46<L^{X-ray} _{bol}~erg~s^{-1}<47}$) we find less clear difference in the distribution of covering factors between Type 1 and Type 2 AGNs ($p=0.25$; right panel of Figure~\ref{fig:KStest}). 

\subsection{The X-ray to MIR relation}

\begin{figure*}
\includegraphics[width=0.5\textwidth]{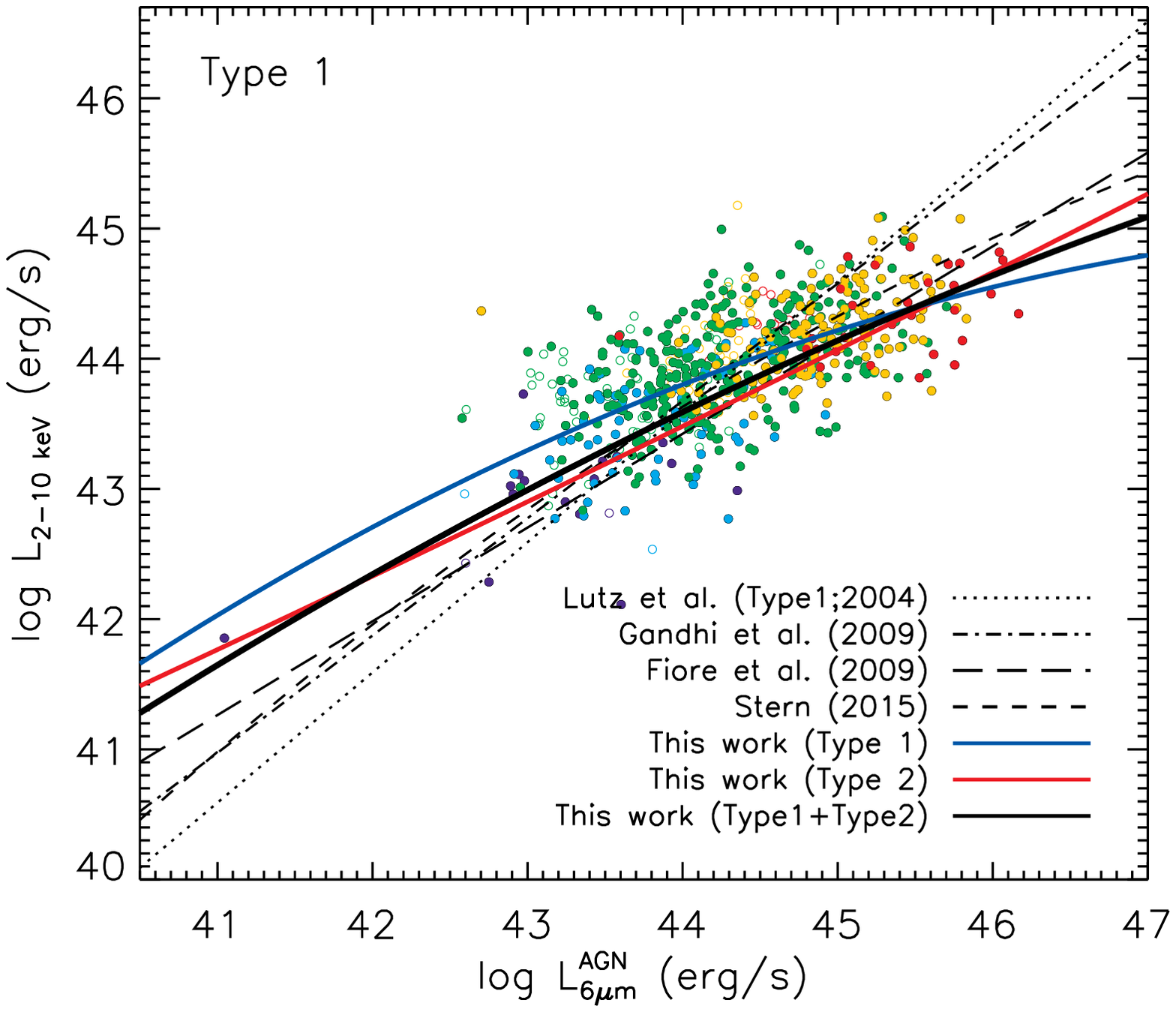}
\includegraphics[width=0.5\textwidth]{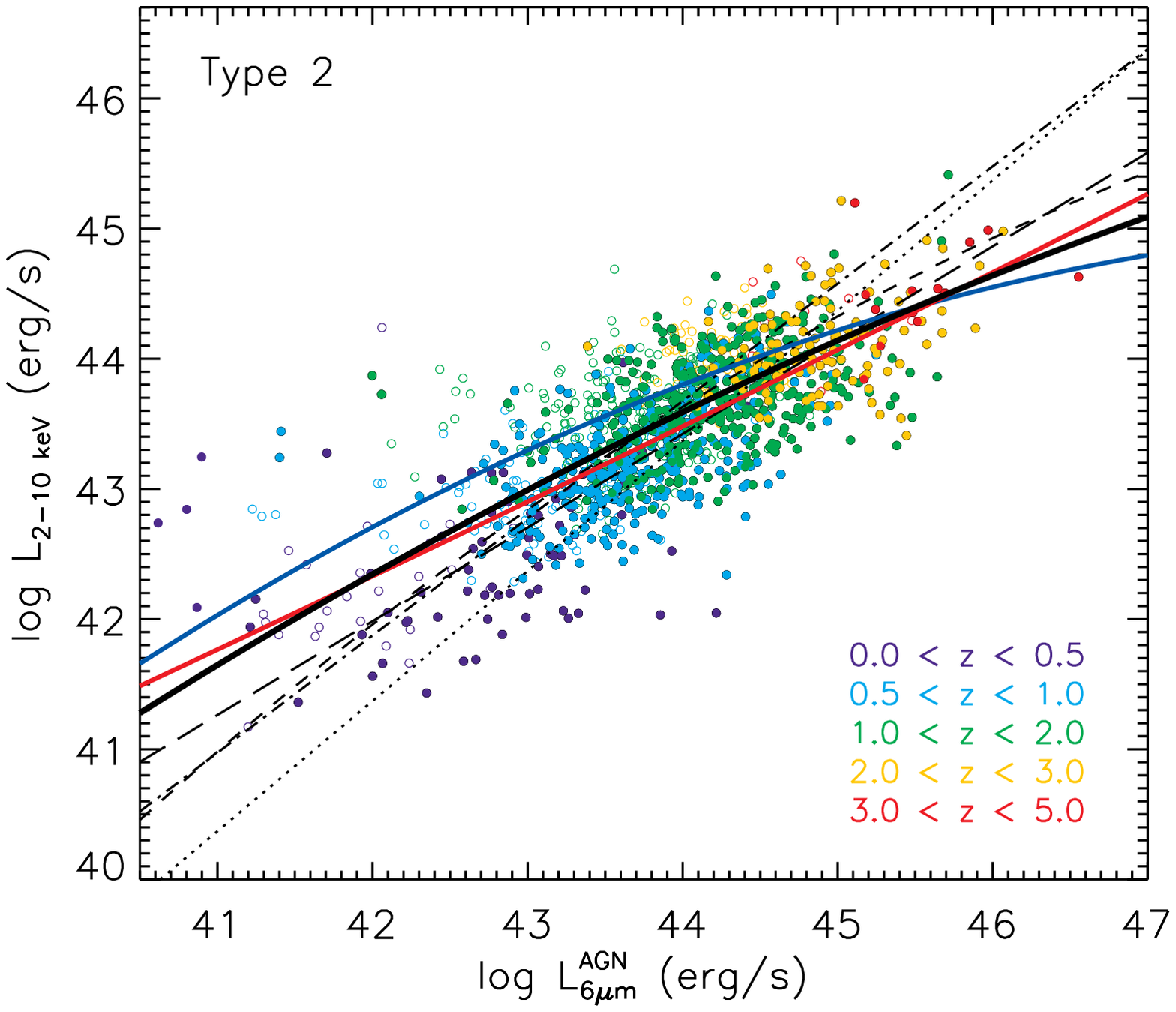}
\includegraphics[width=0.5\textwidth]{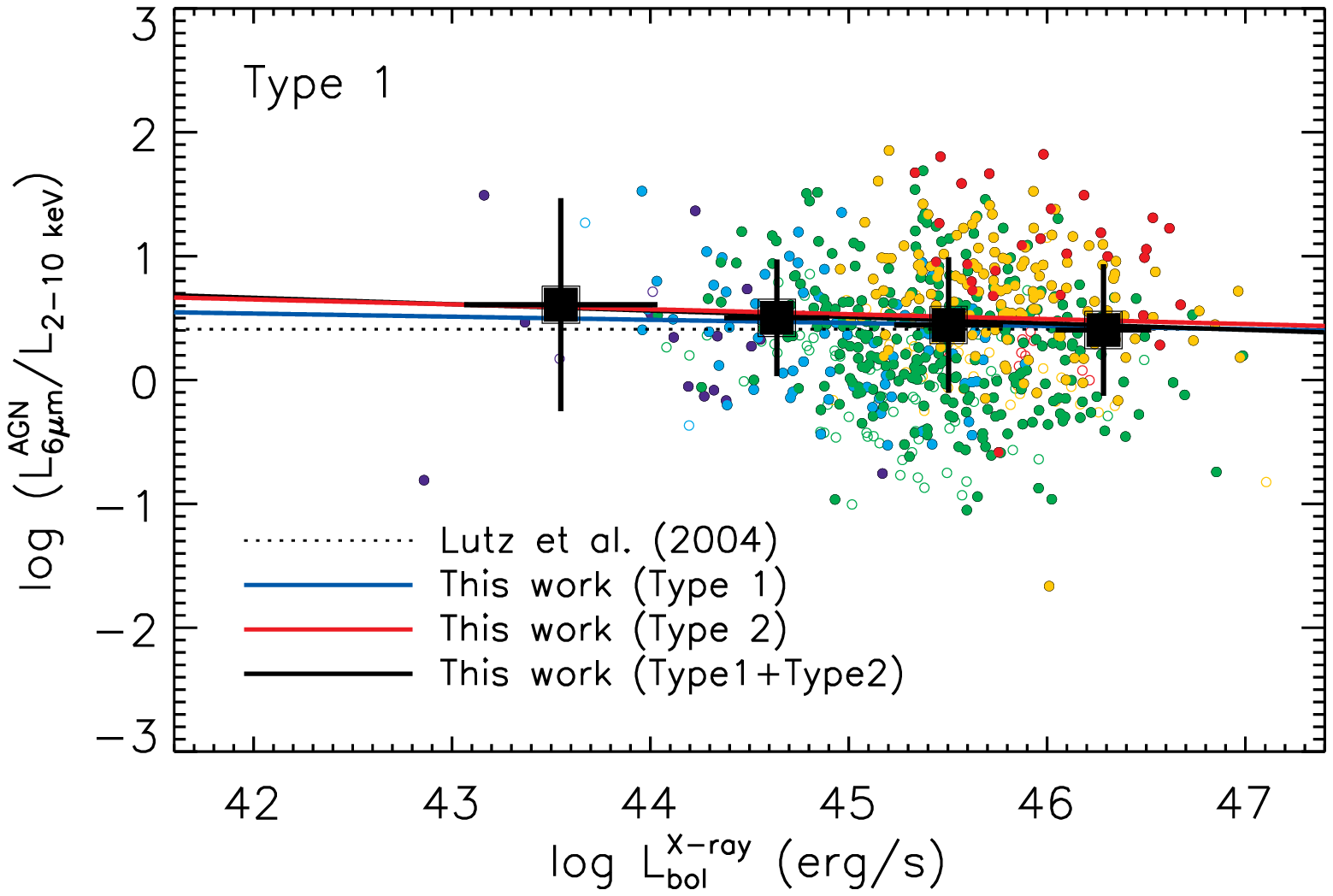}
\includegraphics[width=0.5\textwidth]{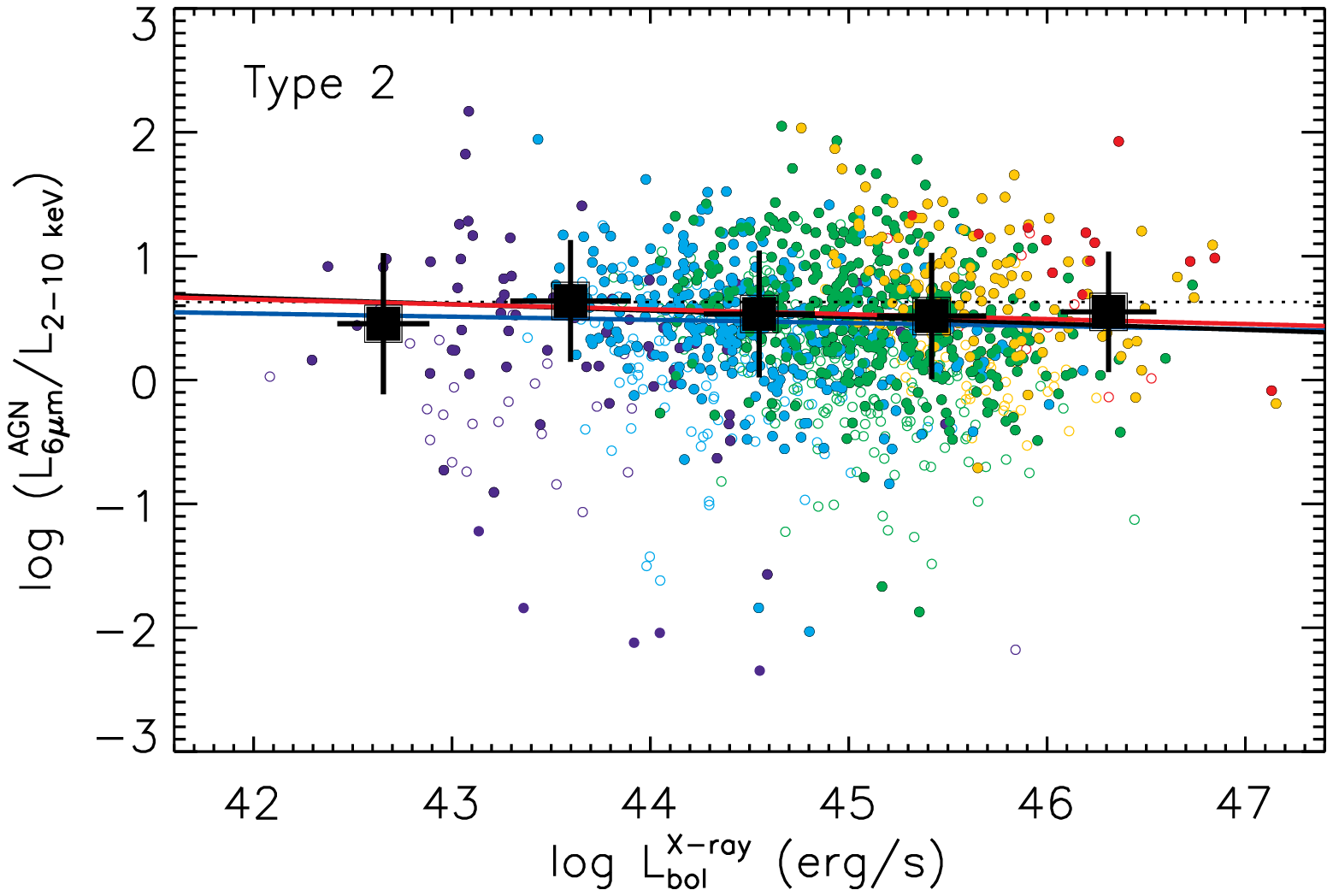}
\caption{Correlation between the intrinsic $L_{2-10~keV}$ and ${\rm L^{AGN} _{6\mu m}}$ for our sample of Type 1 (left) and Type 2 (right) AGNs. Filled symbols indicate {\it Herschel}-detected sources while empty symbols represent {\it Herschel}-undetected sources. The correlation of ${\rm L_{2-10~keV}-L^{AGN} _{6\mu m}}$ for our sample of AGNs is shown as a black solid curve, and those for Type 1 and Type 2 AGNs are indicated as blue and red curves, respectively. For comparison, we also show the relations from \citet[dotted line]{Lutz04}, \citet[dash-dotted line]{Gandhi09}, \citet[long-dashed line]{Fiore09} and \citet[dashed curve]{Stern15}. In the bottom panels, we plot the ratio between $L_{2-10~keV}$ and ${\rm L^{AGN} _{6\mu m}}$ versus AGN bolometric luminosity. Black squares indicate mean values in the AGN bolometric luminosity bins. The horizontal dotted line marks the average L$^{\rm AGN} _{6\mu m}$/L$_{\rm 2-10~keV}$ ratios of local Seyfert galaxies from \citet{Lutz04}.}
\label{fig:LxL6}
\end{figure*}

The X-ray and MIR emission is a key to characterize the nuclear regions of AGNs. We investigate the correlation between X-ray emission and AGN MIR luminosity for both Type 1 and Type 2 AGNs over a wide dynamic range in luminosities and redshifts. We derive the monochromatic luminosity of AGNs at rest-frame 6$\mu$m (L$^{\rm AGN} _{6\mu m}$) from the best-fitting AGN torus template. Figure~\ref{fig:LxL6} shows the absorption-corrected intrinsic 2--10 keV X-ray luminosity ($L_{\rm 2-10~keV}$) against the uncontaminated AGN MIR (${\rm L^{AGN} _{6\mu m}}$) luminosity for Type 1 (left) and Type 2 (right) AGNs in the five redshift bins as labeled. Filled and empty symbols indicate the {\it Herschel}-detected and {\it Herschel}-undetected sources, respectively. In the bottom panels of Figure~\ref{fig:LxL6}, we show the ratio of the ${\rm L^{AGN} _{6\mu m}}$ to $L_{\rm 2-10~keV}$ with respect to the AGN bolometric luminosity (${\rm L^{X-ray} _{bol}}$). Black filled squares indicate mean values of ${\rm L^{AGN} _{6\mu m}}$/$L_{\rm 2-10~keV}$ in the ${\rm L^{X-ray} _{bol}}$ bins. The horizontal dotted line marks the average ${\rm L_{\rm X}-L^{AGN} _{6\mu m}}$ ratio of local Seyfert galaxies from \citet{Lutz04}. 

We find a good agreement between our SED best-fitting solution at L$^{\rm AGN} _{6\mu m}$ and the absorption-corrected $L_{\rm 2-10~keV}$, implying that most of the X-ray emission from the accretion disk is re-emitted in the MIR band. We derive a least-squares polynomial fit to our sample of AGNs (black solid curve) as follows:
\begin{equation*}
log~L_{\rm 2-10~keV} = -0.025~x^{2} + 2.744~x -29.418
\end{equation*}
where $x$ is the monochromatic luminosity of the AGN at rest-frame 6$\mu$m ($log~{\rm L_{\nu}(6\mu m}$) ${\rm erg~s^{-1}}$). For each Type 1 and Type 2 AGN sample, we find the polynomial coefficients of ($x^{2},~x,~x^{0}$) to be (-0.043, 4.260, -60.090) for Type 1 (blue solid curve) and (0.003, 0.242, 25.315) for Type 2 AGNs (red solid curve). For comparison, we also show the ${\rm L_{\rm X}-L^{AGN} _{6\mu m}}$ relations from \citet[dotted line]{Lutz04}, \citet[dash-dotted line]{Gandhi09}, \citet[long-dashed line]{Fiore09} and \citet[dashed curve]{Stern15}. We convert the monochromatic luminosity measured at different wavelengths for these comparison samples (i.e., 5.8 $\mu$m and 12 $\mu$m) to L$_{6\mu m}$ using the AGN template. \citet{Lutz04} and \citet{Gandhi09} presented this relation for local Seyfert galaxies, establishing the correlation at low luminosities, while \citet{Fiore09} and \citet{Stern15} investigated this relation for the most luminous quasars, presenting the relation from the Seyfert regime to the powerful quasar regime. \citet{Stern15} has demonstrated a luminosity-dependent $L_{\rm X}-L_{\rm MIR}$ relation for luminous quasars, reporting that the $L_{\rm X}-L_{\rm MIR}$ fit bends at higher luminosities to lower $L_{\rm X}$/$L_{\rm MIR}$ ratios. 

The (heavily) obscured sources, where the X-ray emission is suppressed, are expected to have weak observed X-ray luminosity compared to the MIR emission. Figure~\ref{fig:LxL6} shows the intrinsic $L_{\rm 2-10~keV}$, which is corrected for absorption derived from the X-ray spectral fitting. Our results indicate that both Type 1 and Type 2 AGNs closely follow the intrinsic $L_{\rm X}-L_{\rm MIR}$ relation. The average values of log~(L$^{\rm AGN} _{6\mu m}$/L$_{\rm 2-10~keV}$) of Type 1 and Type 2 AGNs are 0.47 and 0.52 with a scatter of $\sim$0.5 dex. We find that there is no clear difference between the ${\rm L^{AGN} _{6\mu m}}$/$L_{\rm X}$ correlations for Type 1 and Type 2 AGNs at a given ${\rm L^{X-ray} _{bol}}$. \citet{Gandhi09} also found that the obscured (Type 2) AGNs follow the same correlation as the unobscured (Type 1) AGNs without large offsets or scatter. This implies that the MIR emission is a reasonably good estimator of AGN power for both Type 1 and Type 2 AGNs regardless of their obscuration. This may be particularly useful, for example, in heavily obscured AGNs (i.e., Compton-thick AGNs), which are not detected in the X-ray band due to the high absorption. 

\begin{figure*}
\center
\includegraphics[width=0.49\textwidth]{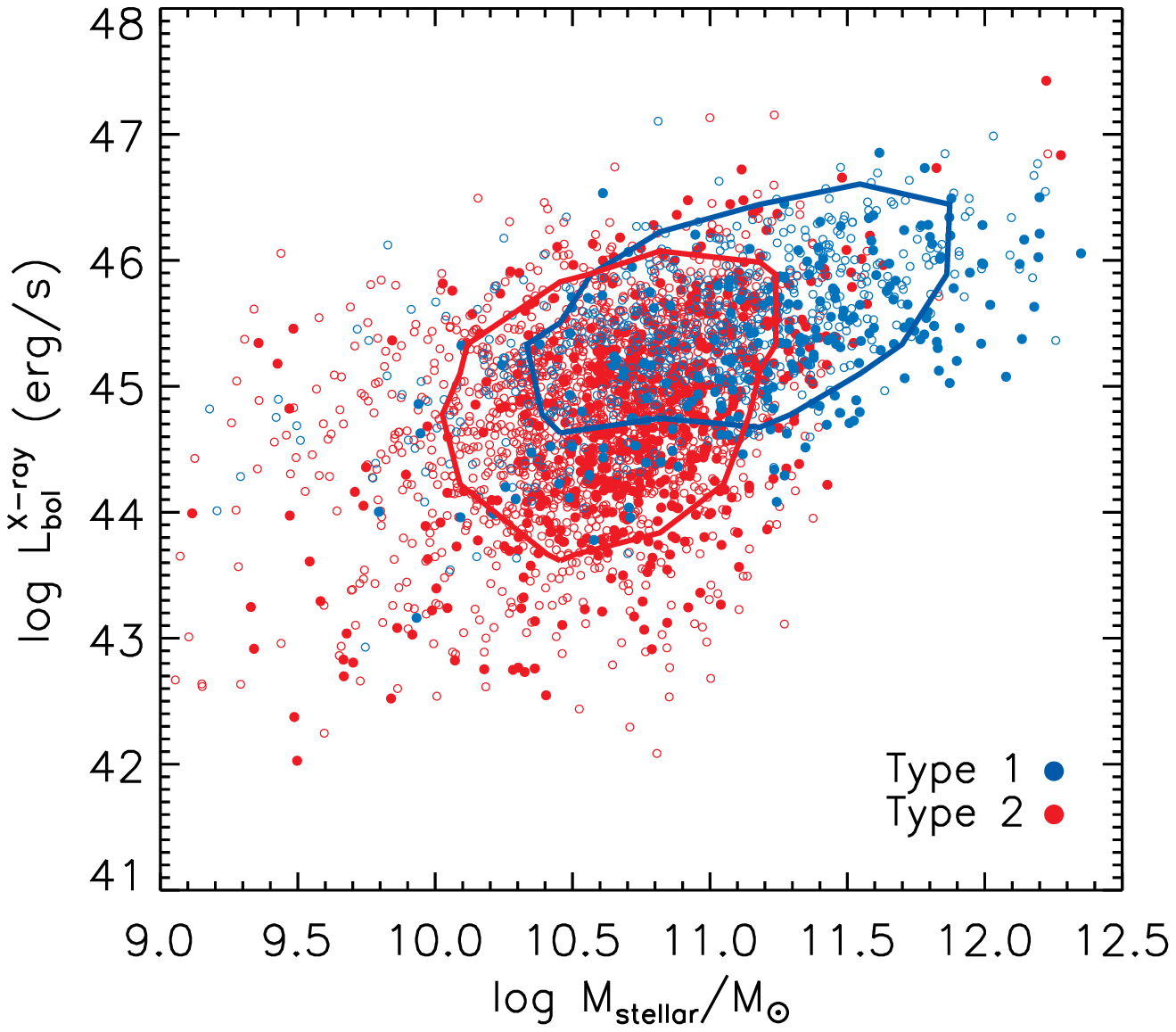}
\includegraphics[width=0.49\textwidth]{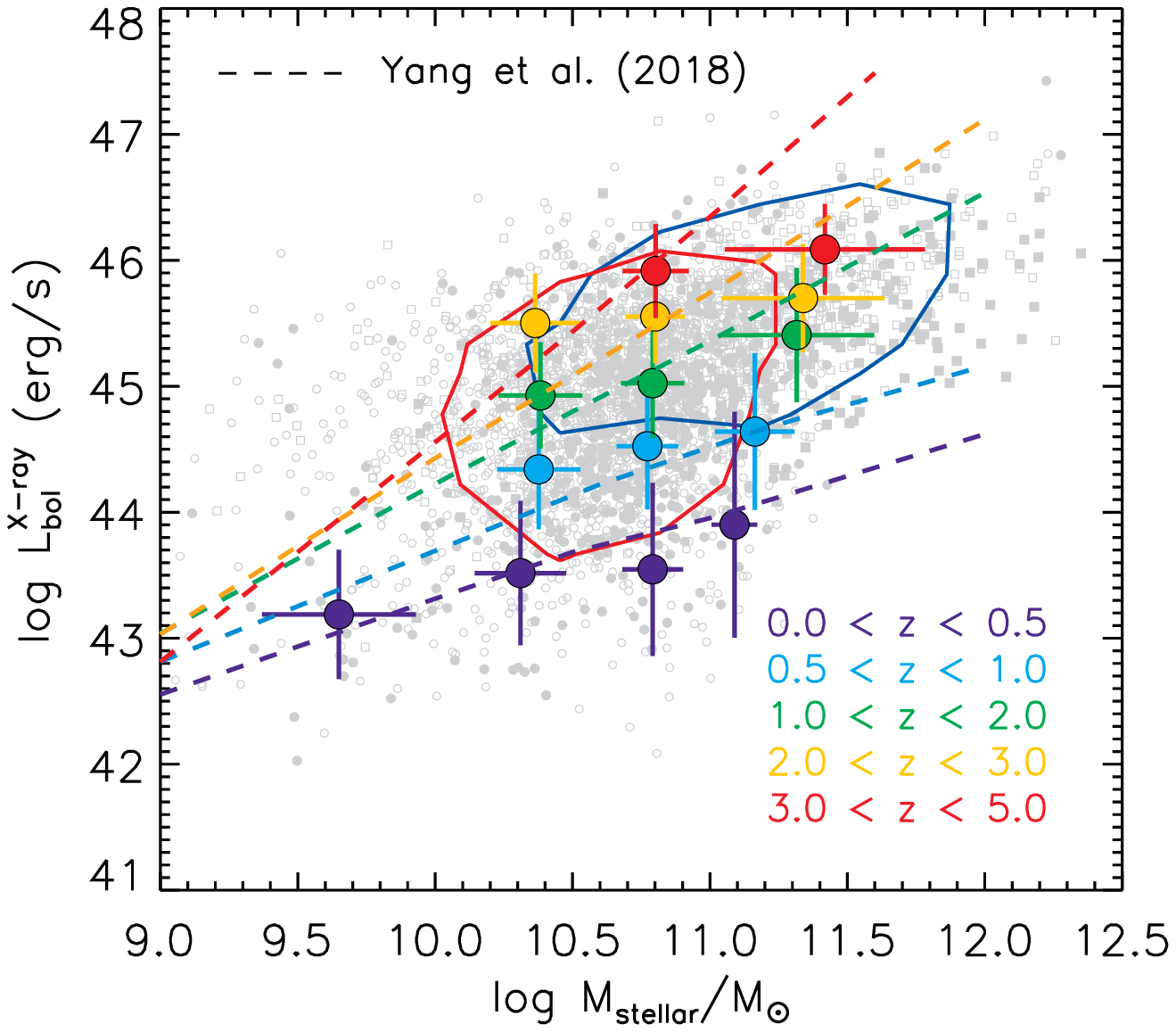}
\caption{AGN bolometric luminosity versus host galaxy stellar mass for Type 1 (blue) and Type 2 (red) AGN host galaxies with contours at the 1$\sigma$ level. Filled symbols indicate MIPS 24$\mu$m detected sources while empty symbols represent 24$\mu$m undetected sources. In the right panel, we show the mean ${\rm L^{X-ray} _{bol}}$ with respect to M$_{\rm stellar}$ in the five redshift bins as labeled. The dashed lines indicate the relationship from \citet{Yang18} for comparison.}
\label{fig:LbolMs}
\end{figure*}

We compute the bolometric correction for MIR luminosity (${\rm L^{AGN} _{6\mu m}}$) using ${\rm L^{X-ray} _{bol}}$, which is derived from the 2--10 keV luminosity with the luminosity-dependent bolometric correction factor described in \citet{Marconi04} (see Section \ref{sec:AGN_lum}). We derive the linear MIR bolometric corrections as follows:
\begin{equation*}
log~L_{\rm bol} = (0.73\pm0.01)~log~\nu L_{\nu}(\rm 6\mu m) + (12.82\pm0.83) \\
\end{equation*}
where ${\rm L_{\nu}(6\mu m}$) is the monochromatic luminosity of AGNs at rest-frame 6$\mu$m from the AGN template. For the Type 1 AGN sample, we find a slope of $0.56\pm0.03$ with a normalization of $20.52\pm1.31$, while for the Type 2 AGN sample, the best-fit slope is $0.74\pm0.02$ with a normalization $12.19\pm1.02$.

\subsection{AGN activity and Stellar mass}

\begin{figure*}
\center
\includegraphics[width=1\textwidth]{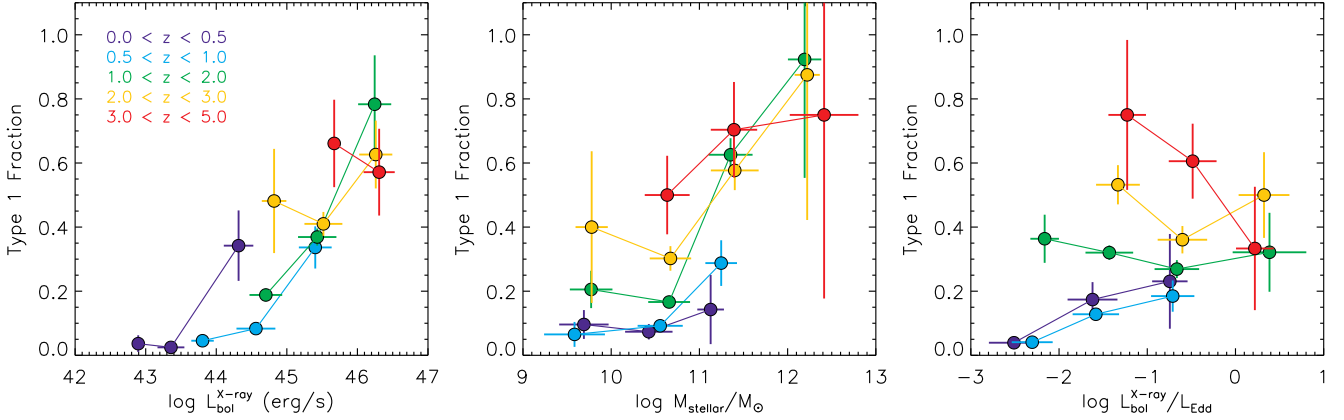}
\caption{The fraction of Type 1 AGNs (the number of Type 1 AGNs divided by the total number of AGNs) with respect to the AGN bolometric luminosity (left), the host galaxy stellar mass (center), and the Eddington ratio (right) in the five redshift bins.}
\label{fig:AGNfrac}	
\end{figure*}

We show the AGN bolometric luminosity versus stellar mass for Type 1 (blue) and Type 2 (red) AGN host galaxies with contours at the 1$\sigma$ level in the left panel of Figure~\ref{fig:LbolMs}. Type 1 and Type 2 AGN host galaxies seem to show significantly different stellar mass distributions based on the K-S test with a confidence level higher than 99.999\% ($p=0.0$). In the right panel, we show the mean ${\rm L^{X-ray} _{bol}}$ with respect to M$_{\rm stellar}$ in the five redshift bins (large colored symbols). We indicate the relationships between the black hole accretion rate (BHAR) and the stellar mass from \citet{Yang18} for comparison (dashed lines). We convert their BHAR, which is derived from X-ray luminosity, to the AGN bolometric luminosity. \citet{Yang18} reported that there is a positive relationship between the long-term SMBH accretion rate and the stellar mass using GOODS-South, GOODS-North, and COSMOS survey data (see also \citealt{Yang17}). They showed that at a given redshift, the BHAR generally increases toward high stellar mass, although the dependence on M$_{\rm stellar}$ is weaker toward the low redshift. While our data at $z<2$ seem to agree fairly well with the relationships of \citet{Yang18}, the Pearson correlation coefficient indicates no clear linear relationship between ${\rm L^{X-ray} _{bol}}$ and M$_{\rm stellar}$ with r=0.44 ($0.0<z<0.5$), 0.24 ($0.5<z<1.0$), and 0.37 ($1.0<z<2.0$). At $z>2$ our data flatten toward the high stellar mass, with the Pearson correlation coefficient r=0.01 ($2.0<z<3.0$) and 0.005 ($3.0<z<5.0$), indicating that no linear correlation is present. As the most massive galaxies (M$_{\rm stellar}$/M$_{\odot}>10^{11.2}$) at $z>2$ tend to host Type 1 AGNs, it is possible that this flattening at the high-mass end is due to the different accretion mechanisms between Type 1 and Type 2 AGNs. 

In Figure~\ref{fig:AGNfrac}, we compute the fraction of Type 1 AGNs (i.e., the number of Type 1 AGNs divided by the total number of AGNs) with respect to the AGN bolometric luminosity (left), the host galaxy stellar mass (center), and the BHAR (right) in the five redshift bins. To get an estimate of the mass accretion rate onto the black hole, we derive the Eddington ratio, ${\rm L_{bol}/L_{Edd}}$, the ratio between the AGN bolometric luminosity and the Eddington luminosity (${\rm L_{Edd}}$). We compute the Eddington luminosity, ${\rm L_{Edd}=1.3\times10^{38}~M_{BH}/M_{\odot}}$, by estimating an approximate black hole mass using the correlation between the black hole mass and the stellar mass found for local AGNs by \citet{Reines15}. 
 
\begin{figure*}
\center
\includegraphics[width=1\textwidth]{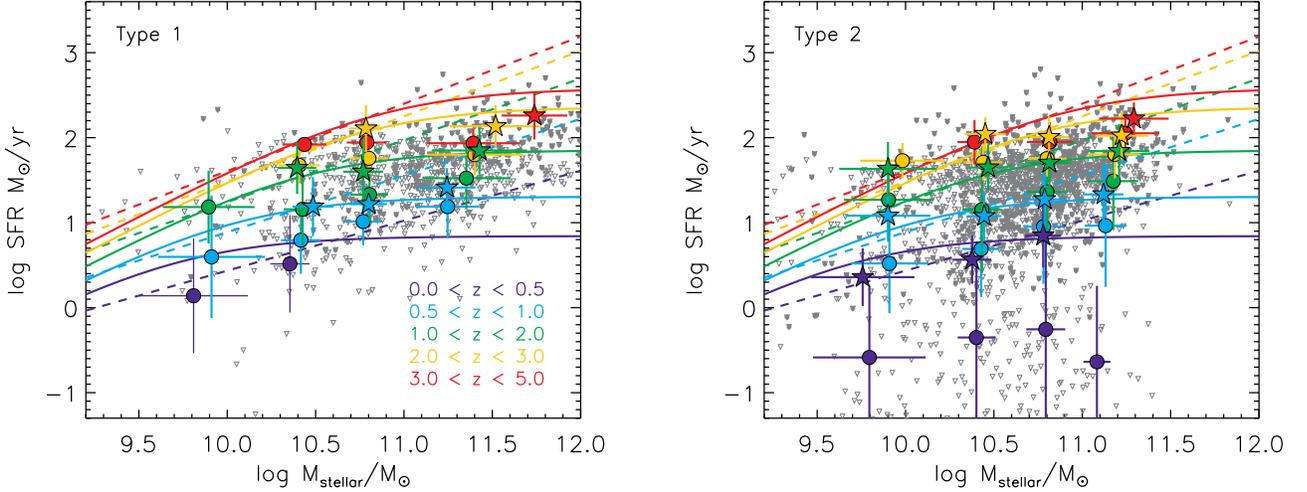}
\caption{SFR versus stellar mass for our sample of Type 1 (left) and Type 2 (right) AGN host galaxies in the five redshift bins. Colored stars represent mean values of SFRs only for {\it Herschel}-detected sources, while colored circles indicate those of SFRs for all of the sources including upper limits of {\it Herschel}-undetected sources. Empty gray circles indicate the individual sources, which are detected in the far-IR {\it Herschel} photometry, and gray downward triangles represent the upper limit SFR for the sources that are not detected in any {\it Herschel} bands. We indicate the star-forming MS relationships from \citet[dashed line]{Speagle14} and \citet[solid curve]{Tomczak16} for comparison. }
\label{fig:SFRMS}
\end{figure*}

We find that there is a strong increase of Type 1 AGN fraction with increasing AGN luminosity, in agreement with previous studies that find a clear decrease in the obscured (Type 2) AGN fraction toward higher AGN luminosity (e.g., \citealt{Simpson05, Hasinger08, Lusso13, Merloni14, Ueda14}). However, we should note that there are underlying correlations among AGN luminosities, BHARs, stellar masses, and redshifts, which could possibly introduce selection biases. Indeed, we also find that there is an increase in Type 1 AGN fraction with increasing host stellar mass in the middle panel in Figure~\ref{fig:AGNfrac}, implying that the luminosity dependence could be a secondary effect in the sense that the AGN activity might be more fundamentally related to host galaxy stellar mass. Recently, \citet{Ricci17} suggested that the strength in radiation pressure from accretion activities is the main driver of observed obscuration fractions, and that Type 1 and Type 2 AGNs are physically different but related accretion mechanisms. We show the dependence on Eddington ratio of Type 1 AGN fraction in the right panel of Figure~\ref{fig:AGNfrac}. While it seems that there is a decline in Type 1 AGN fraction with increasing accretion rates at $z>3$, we do not find a clear dependence of Eddington ratios on different AGN types. On the other hand, the fraction of Type 1 AGNs increases with increasing redshift at a given Eddington ratio. \citet{Treister06} also found that the observed fraction of obscured (Type 2) AGNs declines slightly with redshift, while the intrinsic fraction of obscured (Type 2) AGNs should increase with redshift when correcting for selection biases. While the uncertainties of the stellar mass for Type 1 AGN host galaxies are larger by a factor of $\sim$2 toward lower masses, our result of the dependence of stellar mass on AGN type implies that Type 1 AGNs could be moderate accreting black holes, hosted by more massive galaxies, and as a result be more luminous than Type 2 AGNs. 

\subsection{AGN activity and Star Formation}

\begin{figure*}
\center 
\includegraphics[width=0.95\textwidth]{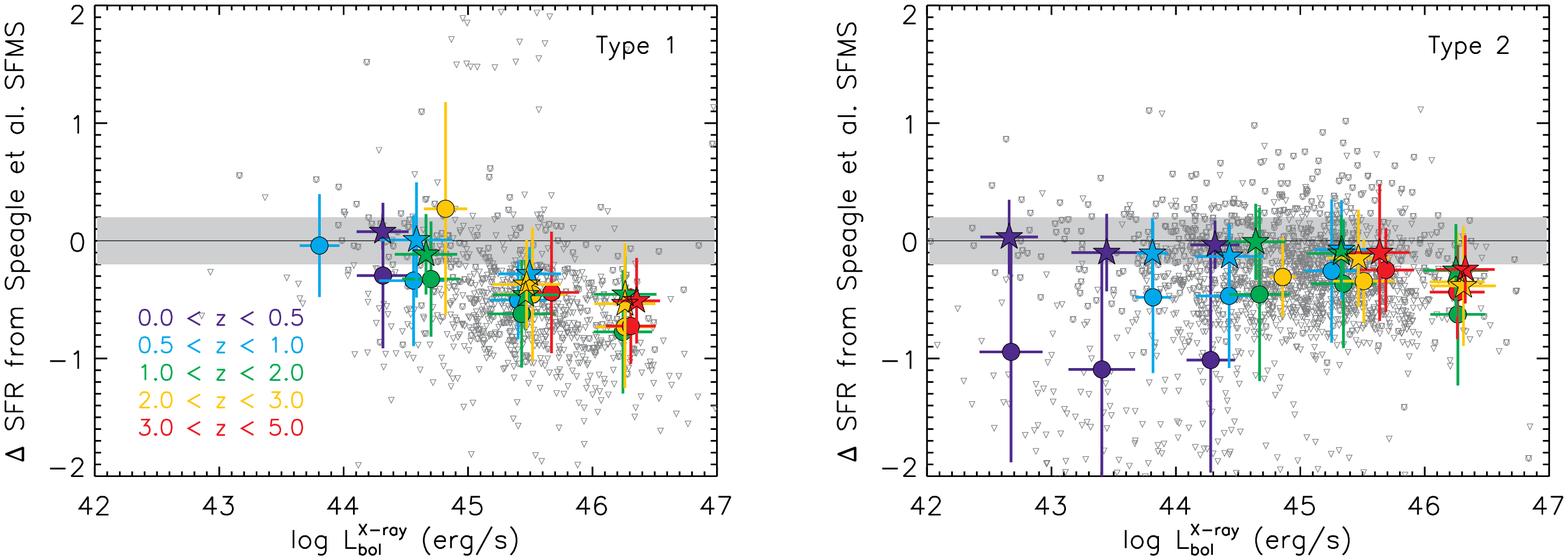}
\includegraphics[width=0.95\textwidth]{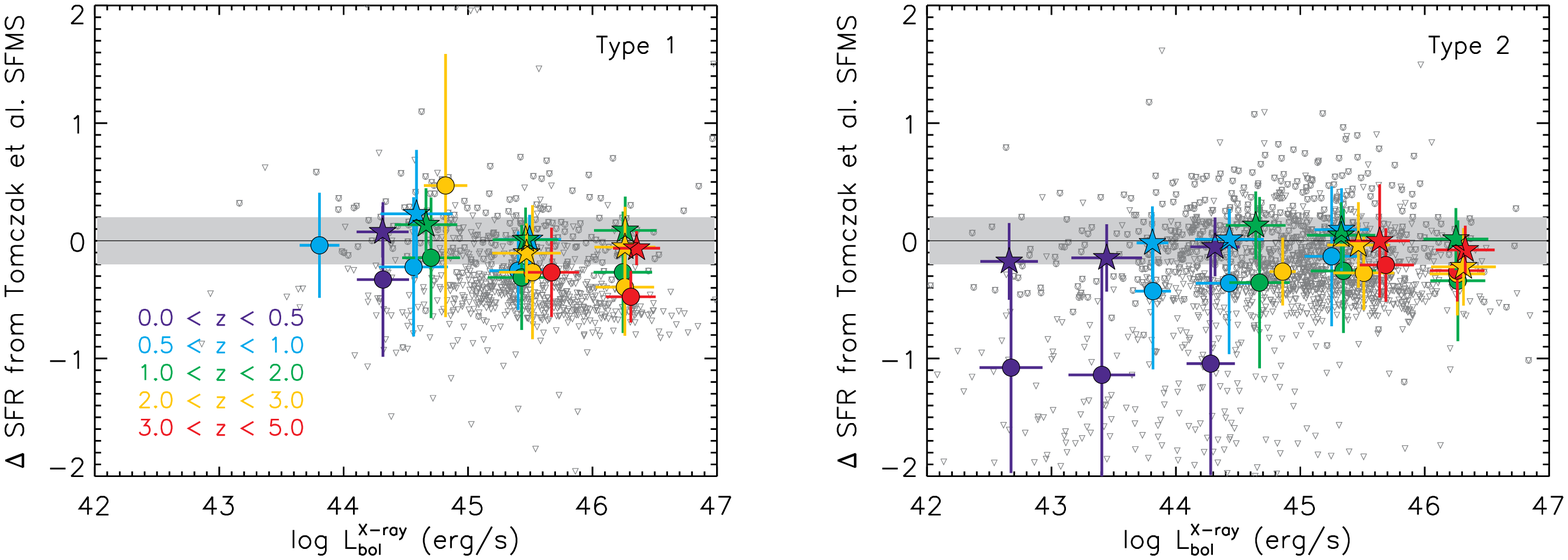}
\caption{SFR offsets ($\Delta$SFR) relative to the star-forming MS relation of  \citet[top; dashed line in Figure~\ref{fig:SFRMS}]{Speagle14} and \citet[bottom; solid line in Figure~\ref{fig:SFRMS}]{Tomczak16} versus AGN bolometric luminosities. The gray shades mark $\Delta$SFR$\sim\pm$0.2 dex. The symbols are same as in Figure~\ref{fig:SFRMS}.}
\label{fig:dSFR}
\end{figure*}

We investigate the impact of AGNs on star formation by analyzing the SFR for AGN host galaxies compared to the main sequence (MS) of star formation (a correlation between SFR and M$_{\rm stellar}$ for normal star-forming galaxies; e.g., \citealt{Daddi07, Elbaz07, Noeske07, Rodighiero11, Speagle14}). We explore the distribution of our sample of AGN host galaxies on the SFR--M$_{\rm stellar}$ diagram in Figure~\ref{fig:SFRMS}. The individual sources are indicated with gray empty circles when the sources are detected in {\it Herschel} FIR photometry, while gray downward triangles represent SFRs that are derived using {\it Herschel} upper limits for the sources detected only up to 24$\mu$m. We show mean SFRs of Type 1 (left) and Type 2 (right) AGN host galaxies for {\it Herschel}-detected sources (colored stars) and mean values for the combination of the SFRs of {\it Herschel}-detected sources and the upper limit SFRs of the {\it Herschel}-undetected sources (colored circles) in the stellar mass bin with uncertainties, split into five redshift bins. We indicate the star-forming MS relationships from \citet[dashed lines]{Speagle14} and \citet[solid curves]{Tomczak16} for comparison. While original star-forming MS studies concluded that the SFR increases with stellar mass as a single power law (dashed lines; see \citealt{Speagle14} for a summary), recent studies have suggested that the star-forming galaxies with low stellar masses (i.e., below $\sim10^{10}$M$_{\odot}$) follow a linear relationship while the SFR--$M_{\rm stellar}$ relation flattens toward the high-mass end (e.g., \citealt{Whitaker14, Lee15, Tomczak16}). In Figure~\ref{fig:dSFR}, we show the SFR offsets ($\Delta$SFR) relative to the star-forming MSs of \citet[linear relationship]{Speagle14} in the top panels and of \citet[M$_{\rm stellar}$-dependent relation]{Tomczak16} in the bottom panels for Type 1 (left) and Type 2 (right) AGN host galaxies with respect to the AGN bolometric luminosities. The symbols are same as in Figure~\ref{fig:SFRMS}. The gray shades mark the intrinsic scatter ($\sim$0.2 dex) of the star-forming MS. 

Overall, AGN host galaxies show significantly broader SFR distributions than star-forming MS galaxies, which is consistent with previous studies (e.g., \citealt{Mullaney15, Shimizu15}). At $z<0.5$, the mean SFRs for Type 2 AGN host galaxies including {\it Herschel}-undetected sources (colored circles) seem to deviate far from the star-forming MS relation, but with large dispersions. Both Type 1 and Type 2 AGN host galaxies with {\it Herschel} detections (colored stars) seem, on average, to have SFRs that lie on the star-forming MS of \citet[solid curves]{Tomczak16} at all redshifts, in good agreement with previous studies (e.g., \citealt{Xue10, Mainieri11, Mullaney12, Rosario13, Suh17}). We find a possible correlation between the SFR offset and the AGN bolometric luminosity for Type 1 AGN host galaxies (top left panel in Figure~\ref{fig:dSFR}), i.e., luminous AGNs tend to have lower SFRs, departing from the linear MS relation (e.g., \citealt{Speagle14}). This could be mainly because Type 1 AGNs tend to be hosted by massive galaxies, and that massive galaxies often have more massive SMBHs (e.g., \citealt{Kormendy13}), which are more capable of accreting gas. We note that when taking into account the dependence of the slope of the star-forming MS on stellar mass \citep{Whitaker14, Lee15, Tomczak16}, AGN host galaxies in the high-mass bins remain on the star-forming MS (\citealt{Tomczak16}; bottom panels in Figure~\ref{fig:dSFR}) over a broad redshift range, and no clear trend is found between the SFR offset and AGN bolometric luminosities for either Type 1 (left) or Type 2 (right) AGN host galaxies, consistent with previous studies (e.g., \citealt{Bongiorno12, Harrison12, Mullaney12, Lanzuisi15, Suh17}). While recent theoretical models have suggested that AGNs are responsible for the flattening of the slope at the highest stellar masses as well as reducing the overall number of massive galaxies (e.g., \citealt{Crain15, Schaye15}), our results indicate that it is not certain whether AGN activities play a role in quenching the star formation or not. Our result implies that the flattening in the star-forming MS at high masses might be primarily related to the host stellar mass.

\begin{figure*}
\center
\includegraphics[width=0.88\textwidth]{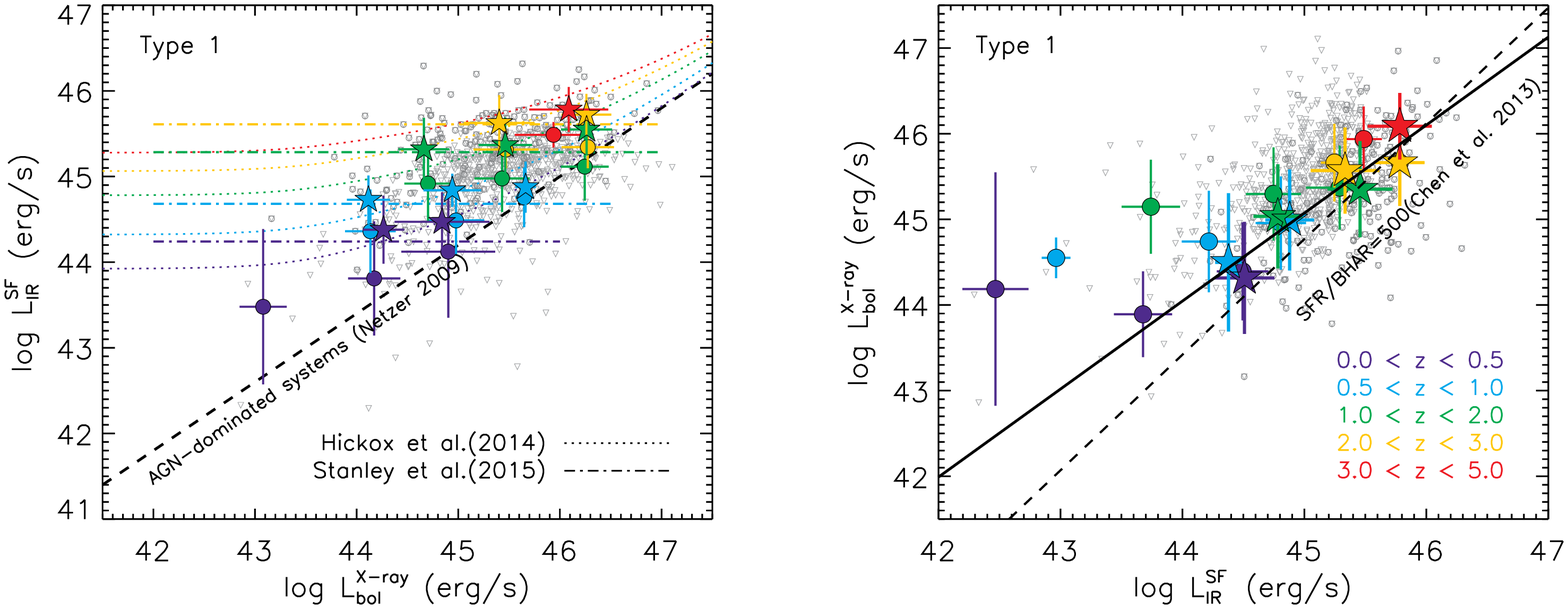}
\includegraphics[width=0.88\textwidth]{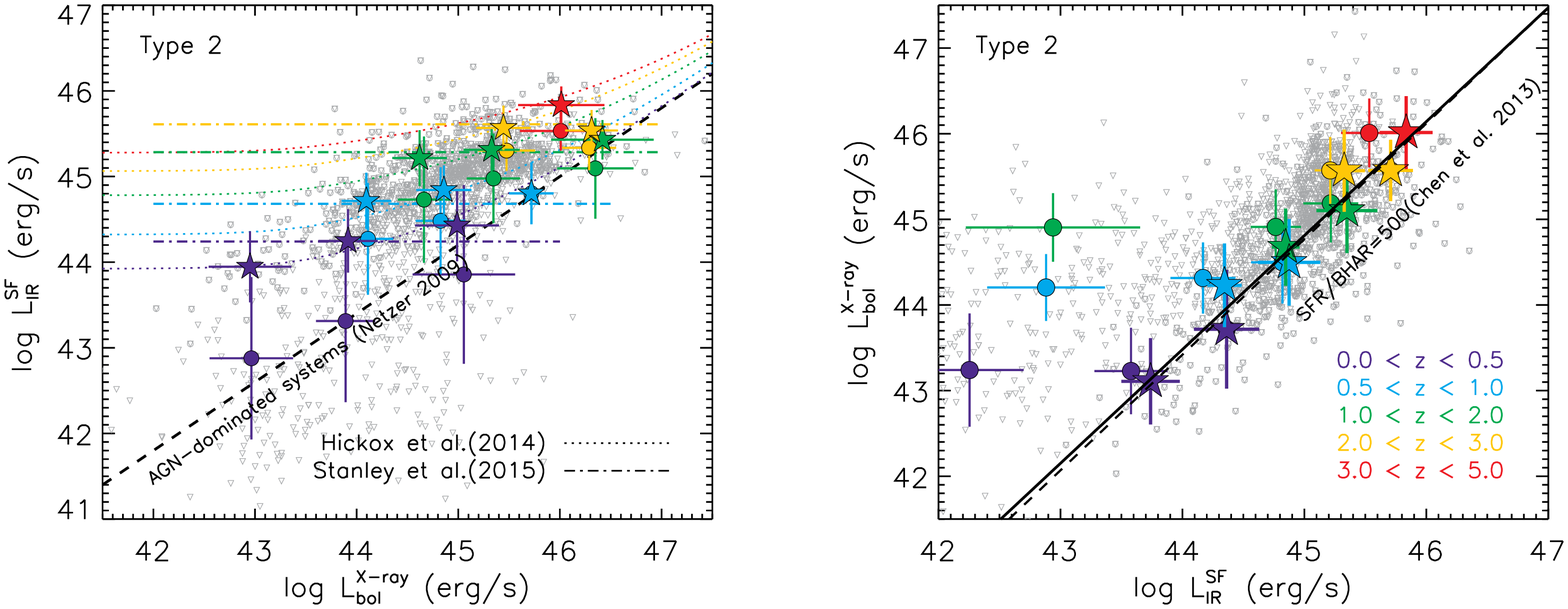}
\caption{Left: distribution of star-forming IR luminosity (${\rm L^{SF}_{IR}}$) versus AGN bolometric luminosity (${\rm L^{X-ray}_{bol}}$) for Type 1 (top) and Type 2 (bottom) AGN host galaxies. The symbols are same as in Figure~\ref{fig:SFRMS}. The colored symbols show the average ${\rm L^{SF}_{IR}}$ in the bins of AGN bolometric luminosity. The black dashed line represents the relationship found in \citet{Netzer09} for objects where IR luminosity is dominated by AGN activity. The colored dashed-dotted lines show the flat relationship in each redshift range from \citet{Stanley15}. The colored dotted curves show the extrapolated trends from \citet{Hickox14} simple model incorporating short-term AGN variability, long-term evolving SFRs, and a universal constant of proportion between SFRs and BHARs. The colors correspond to redshift ranges as labeled. Right: the average ${\rm L^{X-ray}_{bol}}$ in the bins of ${\rm L^{SF}_{IR}}$ for Type 1 (top) and Type 2 (bottom) AGN host galaxies. The solid line indicates the linear fit to our sample of {\it Herschel}-detected sources, and the black dashed line represents the constant proportional relationship between SFR and BHAR found in \citet{Chen13}.}
\label{fig:LsfLbol}
\end{figure*}

We further explore the relationship between star formation and AGN activity of Type 1 (top panels) and Type 2 (bottom panels) AGN host galaxies in Figure~\ref{fig:LsfLbol}. We show the distribution of total star-forming IR luminosity (${\rm L^{SF}_{IR}}$) derived from the best-fitting starburst model, and the AGN bolometric luminosity (${\rm L^{X-ray}_{bol}}$). The symbols are same as in Figure~\ref{fig:SFRMS}. In the left panels of Figure~\ref{fig:LsfLbol}, we show the average ${\rm L^{SF}_{IR}}$ in the bins of ${\rm L^{X-ray}_{bol}}$ for each redshift bin. The colors correspond to redshift ranges as labeled. The black dashed line represents the relationship for objects where IR luminosity is dominated by AGN activity in the local universe presented in \citet{Netzer09}. We show the simple empirical model at each redshift bin from \citet[colored dotted curves]{Hickox14} in which the individual AGNs are allowed to vary on short timescales on the basis of an assumed BHAR distribution, providing the average SFR as a function of AGN luminosity. We also show the flat relationship of ${\rm L^{SF}_{IR}}$ with ${\rm L^{X-ray}_{bol}}$ for each redshift range from \citet{Stanley15} as dash-dotted lines. In the right panels of Figure~\ref{fig:LsfLbol}, we show the average ${\rm L^{X-ray}_{bol}}$ in bins of ${\rm L^{SF}_{IR}}$ in each redshift bin. The dashed line indicates the constant linear relationship between SFR and BHAR found in \citet{Chen13}, and the solid line shows the linear fit to our sample of {\it Herschel}-detected sources.

Across all the individual redshift ranges, we do not find a strong correlation between ${\rm L^{SF} _{IR}}$ and ${\rm L^{X-ray} _{bol}}$, with the Pearson correlation coefficient r$\lesssim$0.2, broadly consistent with the flat relationship suggested by previous studies \citep{Lutz10, Shao10, Harrison12, Mullaney12, Rovilos12, Stanley15, Lanzuisi17}. In the left panels of Figure~\ref{fig:LsfLbol}, we find that our results are in broad agreement with those of \citet[dashed-dotted lines]{Stanley15} with respect to both redshift and ${\rm L^{SF} _{IR}}$. Recent studies have suggested a possible physical explanation for this behavior, i.e., that the shorter variability timescale of AGNs with respect to that of star formation processes could lead to a flat correlation between the SFR and the AGN luminosity when taking the average over the most variable quantity \citep{Hickox14, Volonteri15}. Compared with the model predicted by \citet[dotted curves]{Hickox14}, our results show a flatter trend of ${\rm L^{SF}_{IR}}$ with ${\rm L^{X-ray}_{bol}}$ within each redshift bin, i.e., the ${\rm L^{SF} _{IR}}$ values of the most luminous AGN bin are systematically below the predicted model.

On the other hand, \citet{Chen13} found a ratio of SFR/BHAR$\sim$500 for a sample of FIR-selected AGN host galaxies at $0.25<z<0.8$, suggesting that the global correlation between SFR and BHAR is consistent with a simple picture in which SFR and AGN activity are tightly linked over galaxy evolution timescales. \citet{Lanzuisi17} further confirmed the idea that SMBH accretion and SFRs are correlated, but occur with different variability timescales, by using X-ray-selected AGNs in the COSMOS field. In the right panels of Figure~\ref{fig:LsfLbol}, we find that the average ${\rm L^{X-ray}_{bol}}$ correlates with bins of ${\rm L^{SF}_{IR}}$ when combining all redshifts for {\it Herschel}-detected sources (colored stars) with the Pearson correlation coefficient r=0.95 for Type 1 and 0.96 for Type 2 AGNs, showing a correlation close to SFR/BHAR$\sim$500 found in \citet{Chen13}. Our result for Type 2 AGNs is consistent with \citet{Chen13}, tentatively extending their the results up to $z\sim5$, while Type 1 AGNs show a shallower slope of ${\rm L^{X-ray}_{bol}}$/${\rm L^{SF}_{IR}}$.

However, we should emphasize that only a small fraction ($\sim$30\%) of sources are detected in {\it Herschel} FIR photometry, which could be the most dusty star-forming galaxies with high SFRs, while the majority of AGN host galaxies are faint in the FIR {\it Herschel} photometry. If we take into account the contribution of the {\it Herschel}-undetected sources, we find that the average ${\rm L^{SF} _{IR}}$ drops by $\sim$0.3 dex from that for only {\it Herschel}-detected sources. In the left panels of Figure~\ref{fig:LsfLbol}, almost all of the {\it Herschel}-detected sources lie above the relation computed for AGN-dominated system in \citet{Netzer09}, but the {\it Herschel}-undetected sources might be AGN-dominated systems without any star formation as seen in their upper limit of ${\rm L^{SF}_{IR}}$. Furthermore, when we consider the {\it Herschel}-undetected sources in the right panel (colored circles), the slopes become flatter at each redshift bin than the relation from \citet{Chen13}. Indeed, we have a factor of $\sim$10 deeper X-ray data than the B\"ootes field of \citet{Chen13}. It is clear that our sample of {\it Herschel}-detected sources lie on the relation of SFR/BHAR$\sim$500 \citep{Chen13}, but for the {\it Herschel}-undetected sources we still have AGN activity but suppressed star formation below the {\it Herschel} detection limit. Our results imply that the majority of moderate-luminosity AGNs seem to still be active after the star formation is reduced/quenched, but we cannot simply conclude that the impact of AGNs suppress the star formation, since there is a global correlation between AGN activity and SFR, i.e., a positive trend between ${\rm L^{X-ray} _{bol}}$ and ${\rm L^{SF} _{IR}}$. 

\section{Discussion}

Black holes and galaxies appear to be closely connected, and thus the interaction between black hole accretion and star formation is key to understanding the growth of SMBHs and their host galaxies. It has been well established that AGNs preferentially reside in massive galaxies (e.g., \citealt{Xue10, Lusso11, Bongiorno12, Brandt15}), in good agreement with our result where X-ray-selected AGNs are hosted by more massive galaxies than the average population of galaxies. This could imply that a substantial growth of galaxies has already occurred before black holes reach their final mass. 

We find that there is a clear increase in Type 1 AGN fraction toward higher AGN luminosity, in agreement with previous studies (e.g., \citealt{Simpson05, Hasinger08, Lusso13, Merloni14, Ueda14, Aird15}). This has been interpreted as an intrinsic physical difference in the accretion mechanisms for different AGN types, suggesting that the simplest orientation-based unification scheme needs to be modified to account for the luminosity dependence of the obscuration. We find that there is also an increase of Type 1 AGN fraction toward increasing host stellar mass, implying that the stellar mass might be more fundamentally related to the AGN activity. 

We discuss the star formation in AGN host galaxies, and the relationship between star formation and nuclear activity. Recent studies suggested that the slope of the SFR--M$_{\rm stellar}$ relation (i.e., the MS of star formation) is dependent on stellar mass, such that it appears to flatten at the high-mass end (i.e., M$_{\rm stellar}>10^{10.5-11}$M$_{\odot}$; e.g., \citealt{Whitaker14, Lee15, Schreiber15, Tomczak16}). We show that SFRs of AGN host galaxies are consistent with those flattening star-forming MSs, but with broader dispersions, in a good agreement with previous studies (e.g., \citealt{Xue10, Mainieri11, Mullaney12, Rosario13, Suh17}).

While the majority of AGNs are hosted by massive galaxies, there is no significant difference between the SFRs of AGN host galaxies and those of normal star-forming galaxies when considering the same mass bins. The flattening in the star-forming MS at high masses indicates that massive galaxies have lower specific SFR than less massive galaxies, which could be a consequence of suppressed star formation in massive galaxies. If the SFR reflects the amount of cold gas available, the reduced SFR in massive galaxies indicates that the mass fraction of cold gas drops toward higher stellar mass. The cold gas is also likely responsible for fueling black hole accretion. We find that the average $L^{\rm X-ray} _{\rm bol}$ increases in bins of increasing $L^{SF} _{\rm IR}$ due to star formation, suggesting that there is a close correlation between SFR and the BHARs. 
  
We propose the possible implications for the growth of black holes and galaxies from our study of Type 1 and Type 2 AGN host galaxies. Black holes and galaxies might both have grown predominantly, potentially by major mergers in the early universe ($z>5$). When the galaxy reaches a critical mass (i.e., ${\rm \sim10^{10}~M_{stellar}}$), at which all necessary mass may already exist in galaxies, both star formation and AGN activity slow down due to the lack of cold gas supply. The secular process can trigger a small amount of both star formation and AGN activity, and thus it is likely that relatively massive galaxies grow slowly together with the episodic activity of moderate-luminosity AGNs (i.e., rejuvenation). This is compatible with the presence of AGN host galaxies in the green valley on the color-magnitude diagram (e.g., \citealt{Schawinski10}). This also seems to be consistent with the fact that Type 1 AGN host galaxies are the most massive and their stellar ages derived from SED fitting are similar to those of the red sequence (i.e, old population). If this is the case, then the AGN activity may not suppress or quench the star formation. Our results indicate that stellar mass appears to be the primary factor related to the star formation, as well as the AGN activity. The likely broad physical picture is likely that gas accretion leads to both AGN activity and global star formation over cosmic time, and these could be intimately connected to each other. 
 
\section{Summary and Conclusions}

We present the multi-wavelength properties of one of the largest X-ray-selected samples composed of 3701 AGNs up to $z\sim5$ in the {\it Chandra}-COSMOS Legacy Survey. Leveraging on the extensive multi-wavelength photometric data available in the COSMOS field, we analyze the properties of Type 1 and Type 2 AGN host galaxies by decomposing the AGN emission and their host stellar light using a SED fitting technique. The main results are summarized as follows.

\begin{itemize}
\item[1.] There is a large overlap in the distribution of covering factors (${\rm L_{Torus}/L^{X-ray} _{bol}}$) between Type 1 and Type 2 AGNs, while the majority of Type 1 AGNs are unobscured in X-rays. The AGN MIR luminosity is well correlated with the intrinsic X-ray luminosity for both Type 1 and Type 2 AGNs. Both Type 1 and Type 2 AGNs follow the same ${\rm L_{2-10keV}-L^{AGN} _{6\mu m}}$ correlation regardless of obscuration. 
\item[2.] We found that there is a strong increase in the Type 1 AGN fraction toward higher AGN luminosity. This correlation could possibly be driven by the fact that Type 1 AGNs tend to be hosted by more massive galaxies. Both the AGN luminosity and SFR are consistent with an increase toward high stellar mass, while both relations flatten toward the high-mass end (${\rm M_{stellar}/M_{\odot}}\gtrsim10^{10.5}$), with a correlation coefficient r=0.01, indicating that almost no correlation is present. This flattening at high masses could be interpreted as a consequence of quenching both the star formation and AGN activity in massive galaxies. 
\item[3.] Overall, Type 1 and Type 2 AGN host galaxies seem to have SFRs that lie on the star-forming MS, independent of the AGN luminosity, when taking into account the flattening in the star-forming MS at high masses. This implies that AGN activity does not significantly affect the global star formation in their host galaxies. 
\item[4.] For {\it Herschel}-detected sources, the BHARs and SFRs are correlated up to $z\sim5$. On the other hand, $\sim$73\% of AGN host galaxies in our sample are faint in the FIR (i.e., {\it Herschel}-undetected), implying that the moderate-luminosity AGNs seem to be still active after the star formation is reduced/quenched.
\end{itemize}

Overall, it is not conclusive whether AGN activity plays a role in quenching the star formation in galaxies. We conclude that the stellar mass might be the primary factor related to suppressing both star formation and AGN activity at ${\rm M_{stellar}/M_{\odot}}\gtrsim10^{10.5}$.

\acknowledgments

We thank the anonymous referee for very helpful comments, which helped to improve the quality of the manuscript significantly. E.L. is supported by a European Union COFUND/Durham Junior Research Fellowship (under EU grant agreement no. 609412). M.O. acknowledges support from JSPS KAKENHI grant No. JP17K14257. D.R. acknowledges the support of the Science and Technology Facilities Council (STFC) through grant ST/P000541/1.


\end{document}